\documentclass[letterpaper,twocolumn,amsmath,amssymb,aps,prl,showpacs]{revtex4}
\usepackage[latin1]{inputenc}
\usepackage{color}
\usepackage{amsmath}
\usepackage{amssymb}
\usepackage{graphicx}
\usepackage{esint}

\makeatletter

\pdfpageheight\paperheight
\pdfpagewidth\paperwidth

\@ifundefined{textcolor}{}
{%
 \definecolor{BLACK}{gray}{0}
 \definecolor{WHITE}{gray}{1}
 \definecolor{RED}{rgb}{1,0,0}
 \definecolor{GREEN}{rgb}{0,1,0}
 \definecolor{BLUE}{rgb}{0,0,1}
 \definecolor{CYAN}{cmyk}{1,0,0,0}
 \definecolor{MAGENTA}{cmyk}{0,1,0,0}
 \definecolor{YELLOW}{cmyk}{0,0,1,0}
}

\makeatother

\begin{document}

\title{Extrinsic Spin Hall Effect Induced by Resonant Skew Scattering in
Graphene}

\author{Aires Ferreira$^{1}$, Tatiana G. Rappoport$^{2}$, Miguel A. Cazalilla$^{3,1}$,
and A.~H. Castro Neto$^{1,4}$}

\affiliation{$^{1}$Graphene Research Centre and Department of Physics, National
University of Singapore, 2 Science Drive 3, Singapore 117546, Singapore}

\affiliation{$^{2}$Instituto de Física, Universidade Federal do Rio de Janeiro,
CP 68.528, 21941-972 Rio de Janeiro, RJ, Brazil}

\affiliation{$^{3}$Department of Physics, National Tsing Hua University, and
National Center for Theoretical Sciences (NCTS), Hsinchu City, Taiwan}

\affiliation{$^{4}$Department of Physics, Boston University, 590 Commonwealth
Avenue, Boston, Massachusetts 02215, USA}
\begin{abstract}
We show that the extrinsic spin Hall effect can be engineered in monolayer
graphene by decoration with small doses of adatoms, molecules, or
nanoparticles originating local spin-orbit perturbations. The analysis
of the single impurity scattering problem shows that intrinsic and
Rashba spin-orbit local couplings enhance the spin Hall effect via
skew scattering of charge carriers in the resonant regime. The solution
of the transport equations for a random ensemble of spin-orbit impurities
reveals that giant spin Hall currents are within the reach of the
current state of the art in device fabrication. The spin Hall effect
is robust with respect to thermal fluctuations and disorder averaging. 
\end{abstract}


\maketitle
The spin Hall effect (SHE)~\cite{Dyakonov71,Hirsch99,Zhang00,Jungwirth12},
that is, the appearance of a transverse spin current in a nonmagnetic
conductor by pure electrical control, has been predicted to occur
in materials with large spin-orbit coupling (SOC). Over the last decade,
its study has lead to an intense experimental activity \cite{Kato04,Sih06,Ando12,Valenzuela06,Seki08},
due to its potential application in spintronics. Recently, the SHE
has been explored for replacing ferromagnetic metals with spin injectors
in applications~\cite{Liu11,Liu12}, opening the door to the development
of spintronic devices without magnetic components.

The activation and control of spin-polarized currents is both of fundamental
and technological interest. The SHE could be used for an efficient
conversion of charge current into spin-polarized currents. The ratio
of the spin Hall current to the steady-state charge current, commonly
known as the spin Hall angle $\theta_{\textrm{sH}}$, measures this
efficiency and it is the most important figure of merit for practical
applications. Generally speaking, the SHE in metals and semiconductors
originates from (i) extrinsic mechanisms, which are due to spin-dependent
scattering of charge carriers by impurities in the presence of SOC~\cite{Dyakonov71,Hirsch99,Zhang00},
and (ii) intrinsic mechanisms, entirely due to SOC in the electronic
band structure, which occur in the absence of any scattering process.
In semiconductors, the spin Hall angles are in the range of $0.0001-0.001$~\cite{Kato04,Ando12}.
On the other hand, $\theta_{\textrm{sH}}$ for metals can be considerably
larger, being of the order of $0.01$ for Pt~\cite{Morota11} and
$0.1$ in a recent measurement performed in Ta~\cite{Liu12}.

Since its successful isolation, graphene~\cite{graphene} has also
become the subject of intensive study in spintronics~\cite{Tombos07,Cho07,Han11,Avsar11,Dlubak12}.
In this material, electrons can propagate ballistically and the carrier
density and polarity can be controlled by an external gate. Spin-orbit
and hyperfine interactions are extremely weak in graphene and therefore
the spin coherence length is expected to be long~\cite{Hernando06,Fabian09}.
These characteristics make graphene appealing for passive spintronic
applications, e.g.,~as a high-fidelity channel for spin-encoded information~\cite{Persin12}.
A striking possibility is to modify graphene for active spintronics.
This may be achieved via spin-orbit splitting of the band dispersion,
e.g.,~by bringing heavy metallic atoms in close contact to graphene~\cite{Marchenko12},
or by locally inducing sizeable SOC ($\sim10$~meV)~\cite{ahcn08,weeks11}.
In Ref.~\cite{ahcn08}, distortions induced by covalently bonded
impurities were predicted to produce the desired effect, and \textit{\emph{Ref.~}}{[}\onlinecite{weeks11}{]}
suggests local SOC enhancement via tunneling of electrons in and out
of a heavy atom. Phenomenologically, random spin-orbit fields have
also been predicted to generate nonzero $\theta_{\textrm{sH}}$ \cite{RandomSOC}.
Moreover, it has been proposed that, in the presence of SOC, graphene
could exhibit the quantum spin Hall effect~\cite{Kane06}.

In this Letter, we consider a monolayer of graphene decorated by a
small density of impurities generating a spin-orbit interaction in
their surroundings. We show that a robust SHE develops through asymmetric
(skew) scattering events. Crucially, and unlike two-dimensional electron
gases (2DEGs), for which resonant enhancement of skew scattering~\cite{Raikh08}
requires resorting to fine tuning and sometimes to phenomena such
as the Kondo effect~\cite{Hewson_Book,Nagaosa09}, our proposal takes
advantage of graphene being an atomically thin membrane, whose local
density of states easily resonates with several types of adatoms,
molecules, or nanoparticles. Resonant scatterers have been predicted
to play an important role in charge transport at high electronic densities~\cite{ResScatt,Ferreira11}.
Here, we argue that a similar physics is behind a huge potential of
graphene for the extrinsic SHE. The decoration with small doses of
certain particles only partially suppresses the charge carrier mobilities
of graphene devices, which combined with large spin diffusion lengths
and Fermi energy tunability, makes this material a promising candidate
for spintronic integrated circuits with SHE-based spin-polarized current
activation and control.

According to our calculations, the extrinsic spin Hall effect in graphene,
as that recently reported in hydrogenated graphene samples~\cite{Balakrishnan13},
can originate from skew scattering alone. The latter is absent in
the first Born approximation~\cite{Ballantine} and, therefore, we
compute transport relaxation rates nonperturbatively via exact partial-wave
expansions. \textcolor{black}{Our results indicate that functionalized
graphene can deliver spin Hall angles comparable to those found in
pure metals ($\theta_{\textrm{sH}}\sim0.01-0.1$~\cite{Morota11,Ando12,Kato04}).}

\begin{figure}
\centering{}\includegraphics[clip,width=1\columnwidth]{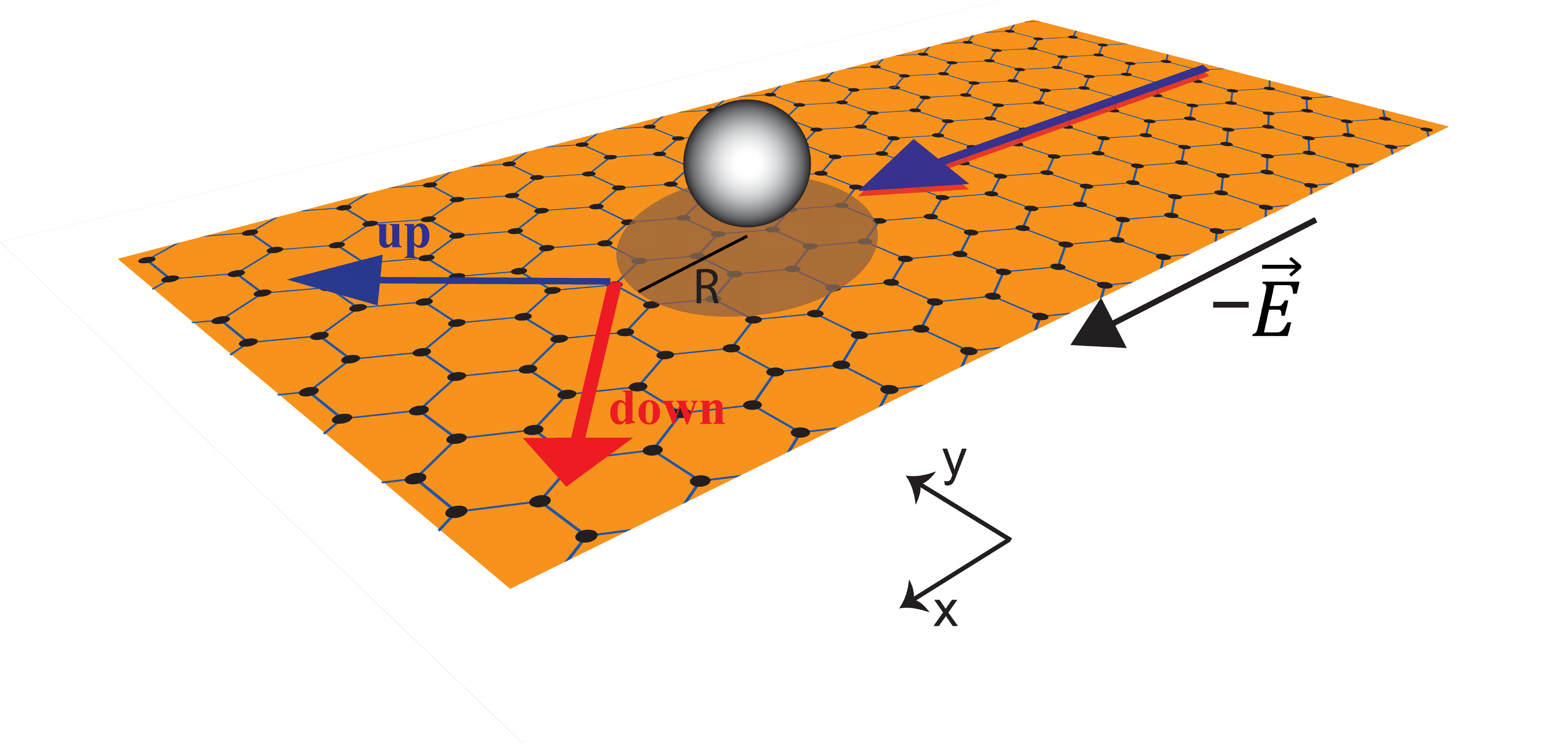}\caption{\label{fig:schematic}Schematic picture of extrinsic spin Hall effect
generated by transport skewness. An impurity (sphere) near the graphene
sheet causes a local spin-orbit field with range $R$. The scattering
of components with positive (negative) angular momentum is enhanced
(suppressed) for charge carriers with $s_{z}=1$ ($s_{z}=-1$), resulting
in a net spin Hall current.}
\end{figure}

In order to investigate the extrinsic SHE and its dependence on Fermi
energy and temperature, we consider a continuum model of graphene
decorated with a small concentration of impurities that locally generate
SOC over nanometer-size regions. The latter could be metallic nanoparticles
inducing SOC via the proximity effect, but other physical realizations
are also possible. (In fact, adatoms in graphene often cluster due
to ripples \cite{Rappoport09} or due to a low adsorption energy~\cite{Kawakami10}.)

Our starting point is the continuum-limit Hamiltonian of graphene
${\cal H}_{0}=\hbar v_{F}(\tau_{z}\sigma_{x}p_{x}+\sigma_{y}p_{y})$,
where $\mathbf{p}=(p_{x},p_{y})$ is the 2D kinematic momentum operator
around one of the two inequivalent Dirac points $K$ and $K^{\prime}$,
$v_{F}\approx10^{6}$~m/s is the Fermi velocity, and $\boldsymbol{\sigma}$
and $\boldsymbol{\tau}$ denote Pauli matrices, with $\sigma_{z}=\pm1$
{[}$\tau_{z}=\pm1${]} describing states on the A(B) sublattice {[}at
$K$($K^{\prime}$){]}. The spin-orbit splitting in the band structure
of pristine graphene is of the order of 10~$\mu$eV and therefore
can be safely neglected~\cite{Fabian09}. The large scatterers considered
here induce sizeable local SOC of the intrinsic-type ${\cal V}_{\textrm{SO}}^{(I)}=\Delta_{I}(\mathbf{r})\tau_{z}\sigma_{z}s_{z}$
and/or Rashba-type ${\cal V}_{\textrm{SO}}^{(R)}=\Delta_{R}(\mathbf{r})(\tau_{z}\sigma_{x}s_{y}-\sigma_{y}s_{x})$;
here, $\mathbf{s}$ are Pauli matrices for spin and $\mathbf{r}=(x,y)$
is the charge carrier position. The dependence of $\mathcal{V}_{\textrm{SO}}^{(I)}$
in the spin and orbital operators is the same as the SOC in flat,
pristine graphene. On the other hand, ${\cal V}_{\textrm{SO}}^{(R)}$
originates in perturbations breaking mirror symmetry about the graphene's
plane (e.g., single-site adsorption). The impurity potentials are
assumed to be smooth on the lattice scale and thus sublattice symmetry
breaking terms (crucial in the single adatom limit \cite{Gmitra13})
are not considered here. For such large scatterers intervalley scattering
is negligible and, in the long wavelength limit, assuming that potentials
have radial symmetry, the scatterer is described by 
\begin{equation}
{\cal V}_{\textrm{ad}}(r)=\mathcal{V}_{\textrm{SO}}(r)+V_{0}(r),\label{SOI}
\end{equation}
where $r=|\mathbf{r}|$, and the (spin-independent) electrostatic
potential $V_{0}(r)$ accounts for extra scalar scattering. Thus,
for $r\gg R$, where $R$ is the range of the potential $\mathcal{V}_{\textrm{ad}}$,
the wave function around the $K$ point reads\begin{widetext} 
\begin{equation}
|\psi_{\lambda,\mathbf{k}}(\mathbf{r})\rangle=\left(\begin{array}{c}
1\\
\lambda
\end{array}\right)e^{ikr\cos\theta}|s\rangle+\frac{f_{\lambda}^{ss}(\theta)}{\sqrt{-ir}}\left(\begin{array}{c}
1\\
\lambda e^{i\theta}
\end{array}\right)e^{ikr}|s\rangle+\frac{f_{\lambda}^{s\bar{s}}(\theta)}{\sqrt{-ir}}\left(\begin{array}{c}
1\\
\lambda e^{i\theta}
\end{array}\right)e^{ikr}|\bar{s}\rangle\ ,\label{wavefunction}
\end{equation}
\end{widetext}where $\lambda=\pm1$ indicates the carrier polarity
with energy $\epsilon=\lambda\hbar v_{F}k$, the ket $|s=\pm\rangle$
describes the orientation of the spin along the $z$ axis, perpendicular
to the graphene plane ($\bar{s}\equiv-s$); $f_{\lambda}^{ss}(\theta)$
and $f_{\lambda}^{s\bar{s}}(\theta)$ are the elastic and inelastic
(``spin-flip'') scattering amplitudes at scattered angle $\theta$,
respectively. The latter is related to the $T$~matrix satisfying
the Lippmann-Schwinger equation $\mathcal{T}(\epsilon)=\mathcal{V}_{\textrm{ad}}+\mathcal{V}_{\textrm{ad}}G_{0}(\epsilon)\mathcal{T}(\epsilon)$,
where $G_{0}(\epsilon)$ is the Green's function $G_{0}(\epsilon)=\left(\epsilon-\mathcal{H}_{0}+\lambda i0^{+}\right)^{-1}$.
Thus, $f_{\lambda}^{ss^{\prime}}(\theta)\equiv f_{\lambda,KK}^{ss^{\prime}}(\theta)$
and $f_{\lambda,\tau\tau^{\prime}}^{ss^{\prime}}(\theta)\propto\langle\lambda\mathbf{k}s\tau|\mathcal{T}(\epsilon)|\lambda\mathbf{p}s^{\prime}\tau^{\prime}\rangle$
with $\tau,\tau^{\prime}=K,K^{\prime}$, $k=|\mathbf{k}|=|\mathbf{p}|$,
and $\theta=\angle(\mathbf{k},\mathbf{p})$.

Let us denote as $\mathcal{F}_{\lambda}(\mathbf{k},\mathbf{p})$ the
$4\times4$ matrix whose elements are $f_{\lambda,\tau\tau^{\prime}}^{ss^{\prime}}(\theta)$
in the spin and valley subspace. The symmetries of the Hamiltonian
$\mathcal{H}(r)=\mathcal{H}_{0}+\mathcal{V}_{\textrm{ad}}(r)$ constrain
the general form of the $4\times4$ matrix $\mathcal{F}_{\lambda}(\mathbf{k},\mathbf{p})$,
which, in general, is a linear combination of the $16$ matrices $s_{\alpha}\tau_{\beta}$
where $\alpha,\beta=0,x,y,z$ (where $\alpha=0$ corresponds to the
unit matrix). However, the assumption of no intervalley scattering
implies that $\mathcal{F}_{\lambda}(\mathbf{k},\mathbf{p})$ commutes
with $\tau_{z}$, which means that $\beta=0,z$. Accounting for the
additional symmetries of $\mathcal{H}(r)$, namely time-reversal plus
$C_{\infty v}\times\{E,C_{2}\}$ (where $E$ is the identity, and
$C_{2}$ is a rotation by $\pi$ about the $z$ axis that also exchanges
the valleys $K$ and $K^{\prime}$) leads to 
\begin{align}
\mathcal{F}_{\lambda}(\mathbf{k},\mathbf{p})=a_{\lambda}s_{0}\tau_{0}+\left(b_{\lambda}s_{z}+c_{\lambda}\mathbf{n}\cdot\mathbf{s}\right)(\hat{\mathbf{k}}\wedge\hat{\mathbf{p}})\tau_{0},\label{eq:symsc}
\end{align}
where $\mathbf{\hat{k}}\wedge\mathbf{\hat{p}}=\sin\theta$ and $\mathbf{n}=\mathbf{\hat{k}}-\mathbf{\hat{p}}$.
The coefficients $a_{\lambda},b_{\lambda},c_{\lambda}$ are complex-valued
functions of $k$ and $\mathbf{\hat{k}}\cdot\mathbf{\hat{p}}=\cos\theta$.
The matrix $\mathcal{F}_{\lambda}(\mathbf{k},\mathbf{p})\propto\tau_{0}$
and therefore valley indices will be suppressed henceforth. Note that,
e.g.,~for scatterers with intrinsic SOC, the component of the spin
perpendicular to the graphene plane ($s_{z}$) is conserved, which
leads to $c_{\lambda}=0$. In general, when the spin-quantization
axis is chosen along the $z$ axis, the terms proportional to $c_{\lambda}$
describe the spin-flip scattering, whereas the term proportional to
$b_{\lambda}$ is responsible for the skew scattering. Equation~(\ref{eq:symsc})
can be used to show that the spin-flip components $\propto c_{\lambda}$
do not contribute to the skew scattering cross section because $|f_{\lambda}^{s\bar{s}}(\theta)|^{2}$
is an even function of $\theta$. This result also applies to the
ensemble of scatterers studied below, for which charge carrier transport
is described by the Boltzmann equation whose collision integral is
determined by the elements of $\mathcal{F}_{\lambda}(\mathbf{k},\mathbf{p})$.

Next, we briefly explain how the spin Hall effect is enhanced by a
single scatterer through the skew scattering mechanism, and the important
role played by resonant scattering in graphene, as well as the main
differences with a 2DEG. To this end, let us consider a scattering
center inducing (locally) an intrinsic SOC, i.e.,~$\Delta_{I}(r)\neq0$.
As noted above, this type of SOC conserves $s_{z}$ and therefore
$c_{\lambda}=f_{\lambda}^{s\bar{s}}(\theta)=0$. The details of the
calculation of $f_{\lambda}^{ss}(\theta)$ and the spin Hall angle
are provided in the Supplemental Material (SM). Here it is sufficient
to realize that, owning to the structure of the extrinsic spin-orbit
coupling term $\Delta_{I}\tau_{z}\sigma_{z}s_{z}$ {[}$(\nabla V_{0}(r)\times\mathbf{p})\cdot\mathbf{s}$
in a 2DEG{]}, SOC induces left-right assymmetry $|f_{\lambda}^{ss}(\theta)|\neq|f_{\lambda}^{ss}(-\theta)|$.
SOC still preserves time-reversal symmetry, which then favors up and
down spins to scatter symmetrically around the incident direction,
i.e., $|f_{\lambda}^{ss}(\theta)|=|f_{\lambda}^{\bar{s}\bar{s}}(-\theta)|$,
thus explaining the formation of a net spin Hall current as depicted
schematically in Fig.~\ref{fig:schematic}. Indeed, at the level
of a single scattering event, the skew cross section 
\begin{equation}
\Sigma_{\perp}^{s}=\int_{0}^{2\pi}d\theta\sin\theta\,|f_{\lambda}^{ss}(\theta)|^{2}\label{eq:skew_transport_section}
\end{equation}
is nonzero and has opposite signs for spins up and down. Finite (nonzero)
$\Sigma_{\perp}^{s}$ is the hallmark of skew scattering. Clearly,
the latter effect is absent in the first Born approximation, according
to which the scattering amplitudes at angles $\pm\theta$ coincide
and hence Eq.~(\ref{eq:skew_transport_section}) is identically zero.
Moreover, we found that, contrary to the case of a 2DEG, a nonperturbative
treatment of the SOC potential $\mathcal{V}_{\textrm{SO}}$ is in
general required and that, in certain cases, the distorted wave Born
approximation, which can be successfully used to treat SOC in the
2DEG~\cite{Ballantine,Raikh08}, fails to describe $\Sigma_{\perp}^{s}$
correctly. A few examples illustrating the perturbative treatments
and a discussion of their limitations in graphene are provided in
the SM.

\begin{figure}
\begin{centering}
\includegraphics[clip,width=1\columnwidth]{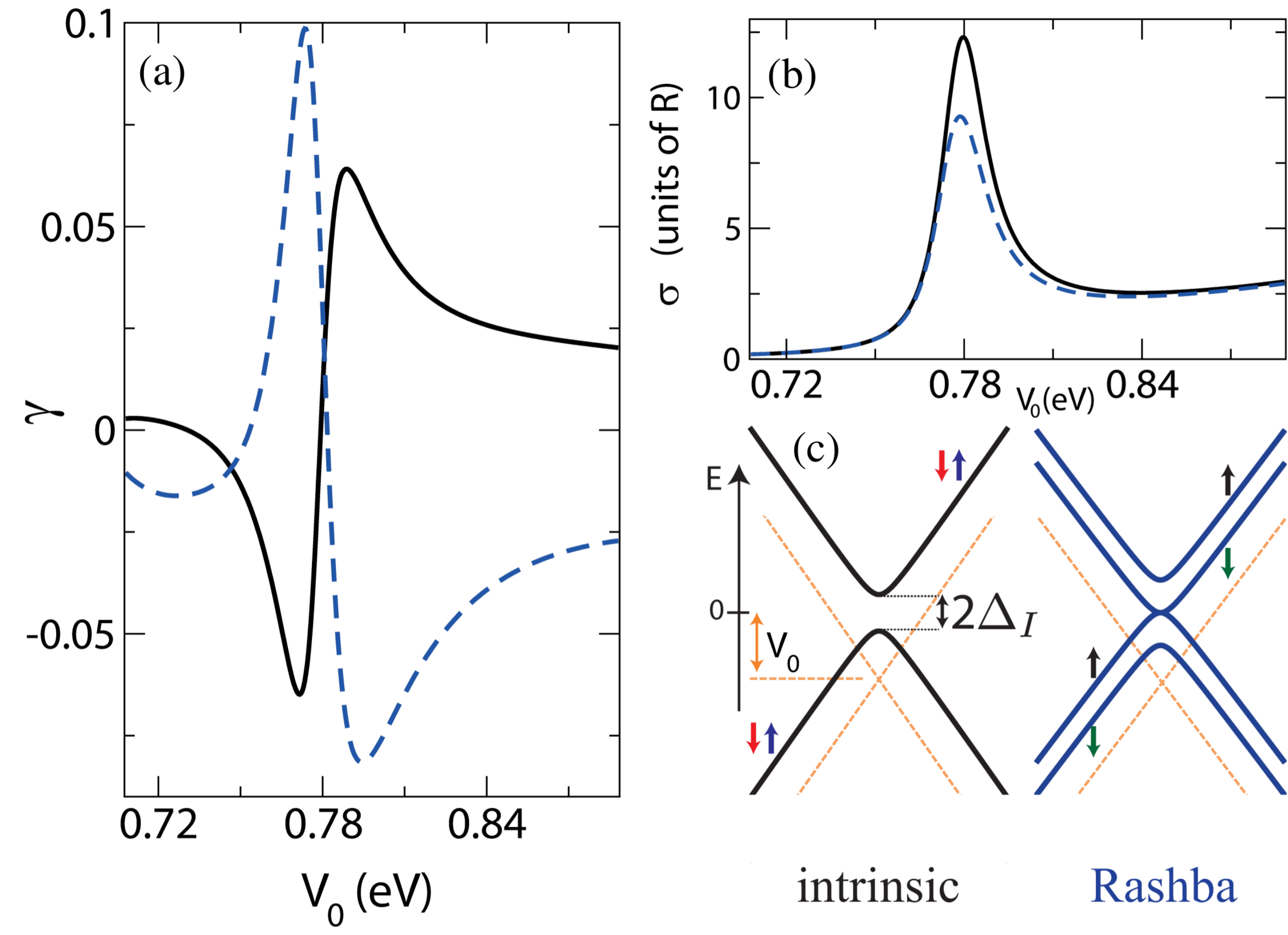} 
\par\end{centering}

\caption{\label{fig:skewness}Skew scattering induced by SOC impurities close
to a resonance in the cross section. (a) Skeweness $\gamma=\Sigma_{\perp}^{s}/\Sigma_{\parallel}^{s}$
as a function of $V_{0}$ for an intrinsic (Rashba)-type SOC scatterer
{[}solid black line (dashed blue line){]}. Even larger values of $\gamma$
are found near sharper resonances occurring at larger $V_{0}$ (not
shown). (b) Transport cross section versus $V_{0}$. These panels
have $R=4$\,nm, $\hbar v_{F}k=0.1$\,eV, and $\Delta=25$\,meV.
(c) Dispersion relation inside the SOC disk scatterer. Dashed orange
lines are guidelines to the eye representing the bulk band structure
of monolayer graphene.}
\end{figure}

As a measure of asymmetry in scattering events we adopt the so-called
transport skewness; for intrinsic SOC scatterers, the latter is defined
as $\gamma\equiv\Sigma_{\perp}^{s}/\Sigma_{\parallel}^{s}$, where
$\Sigma_{\parallel}^{s}=\int d\theta(1-\cos\theta)\,|f_{\lambda}^{ss}(\theta)|^{2}$
is the transport cross section for a carrier with spin $s$ {[}for
Rashba SOC see the discussion below Eq.~(\ref{eq:sH_angle}){]}.
Exact evaluations show (i) $|\gamma|>0$ for local SOCs of the intrinsic
type, (ii) local Rashba SOCs induce $|\gamma|>0$ provided that electron-hole
symmetry is broken by an electrostatic term, i.e.,~$V_{0}\ne0$,
and (iii) $|\gamma|$ is maximum near resonances in $\Sigma_{\parallel}^{s}$.
To illustrate these findings, we model the SOC active impurity as
a uniform disk scatterer of radius $R$ (see Fig.~\ref{fig:schematic}),
according to ${\cal V}_{\textrm{ad}}(r)=[V_{0}+\mathcal{V}_{\textrm{SO}}^{(I/R)}]\Theta(R-r)$,
with $\Theta(.)$ denoting the Heaviside step function and $\mathcal{V}_{\textrm{SO}}^{(I/R)}$
being intrinsic or Rashba-type SOC with $\Delta_{I/R}(\mathbf{r})\equiv\Delta$.
The different symmetries of these terms justifies studying them separately.
Furthermore, it can be shown that interference between intrinsic and
Rashba SOC does not suppress the resonant behavior of skewness (see
SM). In our calculations we have taken $\Delta\sim10$~meV, which
is consistent with \emph{ab initio} calculations for metal atoms adsorbed
in graphene~\cite{weeks11,Niu11}. The skewness of SOC active disk
scatterers in the vicinity of a particular resonance is shown in Fig.~\ref{fig:skewness}.
The function $\gamma(V_{0})$ follows an approximately asymmetric
shape for both intrinsic and Rashba SOC. We further note that for
Rashba-only SOC the skewness approaches zero as $V_{0}\rightarrow0$
(not shown). We also found that $\gamma$ is larger near sharp resonances,
typically occurring at large $V_{0}$. It is known that small doses
of certain adatoms with large effective $V_{0}$ values produce resonances
near the Fermi level of graphene~\cite{Ferreira11} that might dominate
charge transport (see Ref.~{[}\onlinecite{ExperimentsRS}{]} for
transport measurements in graphene covered with hydrogen). For dilute
SOC disorder, the parameter $\gamma$ can therefore be seen as a figure
of merit for the capability of generating net transverse spin currents
via skew scattering. In fact, as shown in what follows, in the absence
of other sources of impurities and at zero temperature, the spin Hall
angle equals $\gamma$. Crucially, the results in Fig.~\ref{fig:skewness}
show that a large $V_{0}$ is not a necessary condition to obtain
large skewness: although resonant impurities such as H induce giant
effective potentials $V_{0}\sim100$\,eV (see Ref.~{[}\onlinecite{Ferreira11}{]}
and the references therein) and significant SOC via lattice distortion~\cite{ahcn08,Gmitra13,Balakrishnan13},
clusters leading to $\mathcal{V}_{\textrm{SO}}$ of tens of mili-electron-volts
most likely produce $V_{0}$ values below those found for chemisorbed
adatoms. Large SOC active scatterers could be formed by the clustering
of physisorbed transition metals inducing significant local enhancement
of SOC, such as Au or In~\cite{weeks11,Marchenko12}.

After analyzing the SHE due to a single scatterer, we next turn to
the experimentally relevant situation of a dilute random ensemble
of scatterers. We focus on the spin Hall current polarized out of
the plane; see the SM for a discussion of in-plane polarization. Our
goal is to compute the spin Hall angle defined as $\theta_{\textrm{sH}}=j_{\textrm{sH}}/j_{x}$,
with $j_{x}=\sum_{s=\pm}\mathbf{j}_{s}\cdot\mathbf{e}_{x}$ and $j_{\textrm{sH}}=\sum_{s=\pm}s\mathbf{j}_{s}\cdot\mathbf{e}_{y}$
being the expectation values of the (charge) longitudinal and (spin)
Hall currents, respectively. We safely neglect the quantum side-jump
contribution to $j_{\textrm{sH}}$ which is subdominant with respect
to skew scattering in the dilute regime of interest here~\cite{Sinitsyn}.
Semiclassicaly, the current is computed according to $\mathbf{j}_{s}=-eg_{v}\sum_{\mathbf{k}}\delta n_{s}(\mathbf{k})\mathbf{v}_{\mathbf{k}}$,
where $\mathbf{v}_{\mathbf{k}}=(1/\hbar)\nabla_{\mathbf{k}}\epsilon_{\mathbf{k}}$
is the band velocity and $\delta n_{s}(\mathbf{k})=n_{s}(\mathbf{k})-n^{0}(\mathbf{k})$
denotes the deviation of the spin-dependent distribution function
from its equilibrium value $n^{0}(\mathbf{k})$ ($g_{v}=2$ is graphene's
valley degeneracy factor). To describe this situation, we need to
solve the Boltzmann transport equation (BTE), which for the steady
state in the presence of a uniform electric field $\boldsymbol{\boldsymbol{\mathcal{E}}}=\mathcal{E}\mathbf{e}_{x}$
reads as~\cite{BTE_comment} 
\begin{equation}
\nabla_{\mathbf{k}}n_{s}(\mathbf{k})\cdot(-e\boldsymbol{\boldsymbol{\mathcal{E}}})=\sum_{\mathbf{p},s^{\prime}}\left[n_{s^{\prime}}(\mathbf{p})-n_{s}(\mathbf{k})\right]W_{s^{\prime}s}(\mathbf{p},\mathbf{k})\,,\label{eq:BTEs}
\end{equation}
where $W_{ss^{\prime}}(\mathbf{k},\mathbf{k}^{\prime})\propto|f^{ss^{\prime}}(\theta)|^{2}\:\delta(\epsilon_{\mathbf{k}}-\epsilon_{\mathbf{k}^{\prime}})$
with $\theta=\angle\left(\mathbf{k},\mathbf{k}^{\prime}\right)$ is
the quantum-mechanical rate for processes with $\mathbf{k}\rightarrow\mathbf{\mathbf{k}^{\prime}}$
and $s\rightarrow s^{\prime}$. Notice that skew scattering implies
that $W_{ss^{\prime}}(\mathbf{k},\mathbf{k}^{\prime})\neq W_{ss^{\prime}}(\mathbf{k}^{\prime},\mathbf{k})$;
cf.,~Eq.~\eqref{eq:symsc}. Here, $W_{ss^{\prime}}(\mathbf{k},\mathbf{k}^{\prime})\equiv\sum_{\alpha=1}^{R}W_{ss^{\prime}}^{(\alpha)}(\mathbf{k},\mathbf{k}^{\prime})$
takes into account all disorder sources, where $R\ge1$ is the number
of such sources. In linear response, the above BTE admits the following
general solution 
\begin{equation}
\delta n_{s}(\mathbf{k})=\nabla_{\mathbf{k}}n^{0}(\mathbf{k})\cdot\left[A_{s}(\mathbf{k})e\boldsymbol{\boldsymbol{\mathcal{E}}}+B_{s}(\mathbf{k})\left(\mathbf{\hat{z}}\times e\boldsymbol{\boldsymbol{\mathcal{E}}}\right)\right]\,,\label{eq:ansatz}
\end{equation}
where $n^{0}(\mathbf{k})$ is the Fermi-Dirac distribution. With these
definitions, and at zero temperature, one finds $\theta_{\textrm{sH}}=B_{\uparrow}(k_{F})/A_{\uparrow}(k_{F})$,
where $k_{F}$ is the Fermi momentum. The latter expression can be
evaluated in closed form: 
\begin{equation}
\left.\theta_{\textrm{sH}}\right|_{T=0}=\frac{\tau_{\parallel}^{*}(k_{F})}{\tau_{\perp}^{*}(k_{F})}=\bar{\gamma}\,,\label{eq:sH_angle}
\end{equation}
where $\tau_{\parallel}^{*-1}=\sum_{s^{\prime},\mathbf{p}}\left(1-ss^{\prime}\cos\theta\right)W_{ss^{\prime}}(\mathbf{k},\mathbf{p})$
and $\tau_{\perp}^{*-1}=\sum_{s^{\prime},\mathbf{p}}ss^{\prime}\sin\theta\, W_{ss^{\prime}}(\mathbf{k},\mathbf{p})$.
The spin Hall angle $\theta_{\textrm{sH}}$ equals the weighted skewness
as defined by $\bar{\gamma}=\bar{\Sigma}_{\perp}^{*}/\bar{\Sigma}_{\parallel}^{*}$,
where $\bar{\Sigma}_{\parallel(\perp)}^{*}\equiv\sum_{\alpha}\frac{n_{\alpha}}{n}\Sigma_{\parallel(\perp)\alpha}^{*}=(nv_{F}\tau_{\parallel(\perp)}^{*})^{-1}$
and $n=\sum_{\alpha}n_{\alpha}$ is the total areal density of impurities.
The explicit solutions for $A_{s}(B_{s})$ further contain the familiar
scattering times $\tau_{\parallel}$ and $\tau_{\perp}$ that do not
enter in the ratio $B_{s}/A_{s}$. The spin-flip contribution to ``star''
rates differ from standard definitions, e.g.,~$\tau_{\parallel,\textrm{flip}}^{*-1}\sim\int d\theta\left(1+\cos\theta\right)W_{s\bar{s}}(\theta)\neq\tau_{\parallel,\textrm{flip}}^{-1}$.
(For this reason, in the calculation of the skewness of a Rashba scatterer
in Fig.~\ref{fig:skewness} we have used $\Sigma_{\parallel}\rightarrow\Sigma_{\parallel}^{*}=\sum_{s^{\prime}}\int d\theta(1-ss^{\prime}\cos\theta)|f^{ss^{\prime}}(\theta)|^{2}$\textcolor{black}{.})
This fact has been largely unnoticed, which we believe is a consequence
of inadequate treatments of the BTE; relaxation rates found here,
on the other hand, result from the exact solution of linearized BTEs
(see the SM for further details). 

\begin{figure}
\begin{centering}
\includegraphics[width=1\columnwidth]{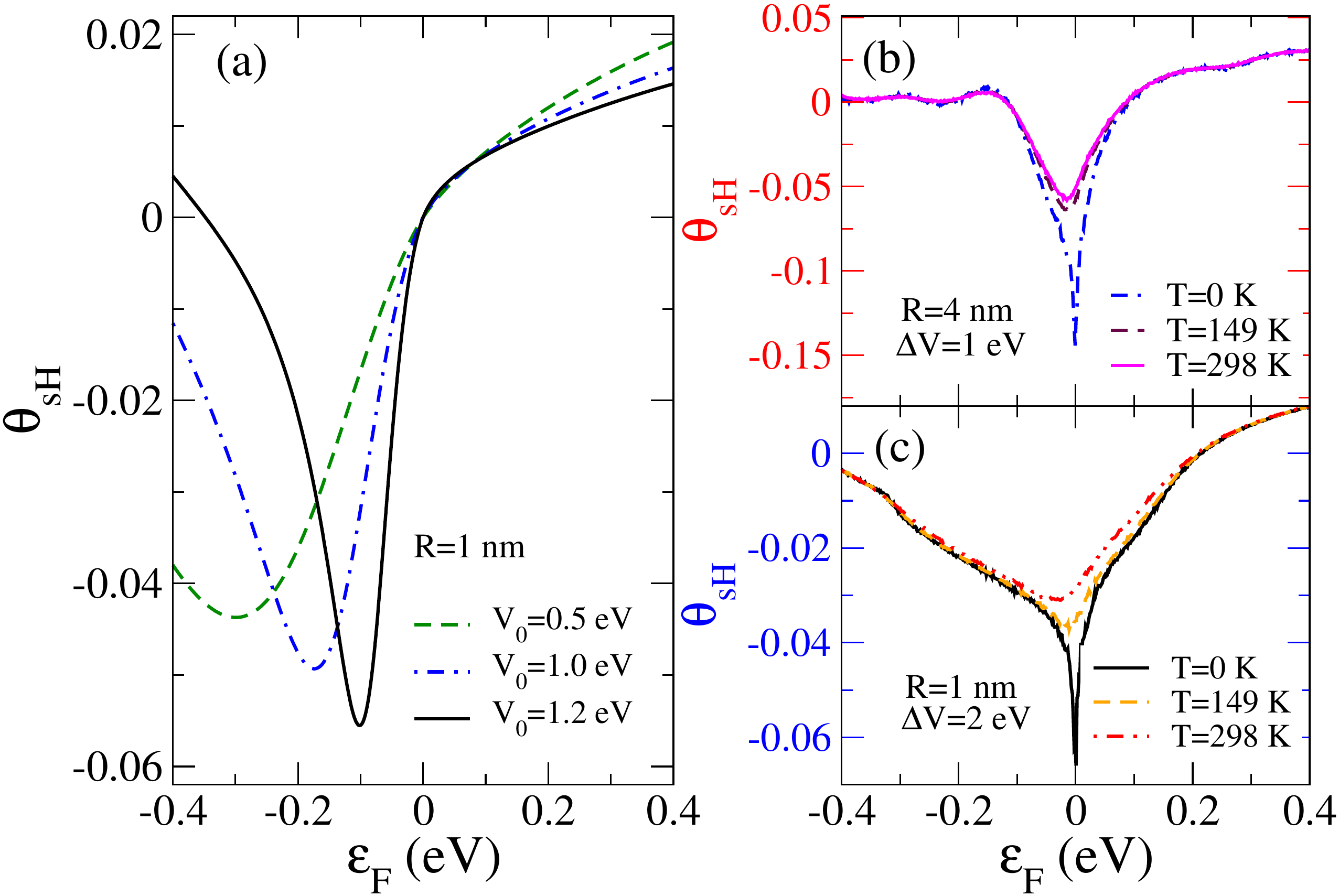}
\par\end{centering}

\caption{\label{fig:roomTangle}Spin Hall angle as a function of Fermi energy
for a dilute random distribution of intrinsic SOC scatterers. (a)
$\theta_{\textrm{sH}}$ at zero temperature for impurities producing
a local electrostatic potential $V_{0}$. (b), (c) $\theta_{\textrm{sH}}$
at different temperatures and considering a random $V_{0}$ potential
with uniform distribution $V_{0}\in[0,\Delta V]$. In all panels we
have taken $\Delta_{I}=25$~meV.}
\end{figure}

A sizeable SHE is expected in relatively clean samples when cross
sections for SOC active scatterers yield the dominant contribution
to both transport and skew cross sections; Fig.~\ref{fig:roomTangle}
shows $\theta_{\textrm{sH}}$ {[}Eq.~(\ref{eq:sH_angle}){]} as a
function of Fermi energy for pristine graphene decorated with a dilute
concentration of intrinsic-type SOC scatterers ($\theta_{\textrm{sH}}$
induced by Rashba-type SOC is of the same order of magnitude and hence
is not shown). The values obtained are comparable with those found
in pure metals \textcolor{black}{$|\theta_{\textrm{sH}}|\sim0.01-0.1$~\cite{Morota11,Liu12}
and are robust with respect to }thermal fluctuations\textcolor{black}{{}
and disorder averaging }{[}compare curves in Figs.~3(a) and 3(c){]};
room temperature spin Hall angles of the order of $0.1$ are obtained
for large scatterers with effective radius of just a few nanometers
{[}see Fig.~3(b){]}. Statistical distribution of scatterer sizes
does not modify qualitatively this picture, indicating that large
SOC active scatterers in clean graphene samples will drive the formation
of robust spin Hall currents. Finally, we verified that time-reversal
symmetry breaking by localized magnetic moments \cite{Lundeberg13}
sitting at the impurities does not suppress the SHE (see the SM).
Our findings suggest that functionalized graphene can be used to design
spintronic integrated circuits with SHE-based spin-polarized current
activation and control.

\emph{Acknowledgements}. A.F., M.A.C. and A.H.C.N. acknowledge support
from the National Research Foundation\textendash{}Competitive Research
Programme through the grant ``Novel 2D materials with tailored properties:
beyond graphene'' (Grant No. R-144-000-295-281). T.G.R. acknowledges
support from INCT\textendash{}Nanocarbono, CNPq, and FAPERJ. M.A.C.
acknowledges support from NSC and start-up funding from NTHU (Taiwan).
Discussions with N.M.R. Peres and A. Pachoud are gratefully acknowledged. 

\begin{center}
\textbf{\large{Supplementary Material}}{\large{{} \setcounter{equation}{0}\setcounter{figure}{0}}} 
\par\end{center}

\section{Analytical Solution of Boltzmann Transport Equations\label{sec:BTE}}

In this section we solve analytically the Boltzmann transport equation
(BTE). The low-energy Hamiltonian is given by $\mathcal{H}=\mathcal{H}_{0}+\mathcal{V}$,
where $\mathcal{H}_{0}$ is the graphene-only term and $\mathcal{V}(\mathbf{r})=\sum_{i}\mathcal{V}^{(i)}(\mathbf{r}-\mathbf{r}_{i})$
is the disordered (spin-orbit) potential due to impurities located
at random positions \{$\mathbf{r}_{i}$\}. Intervalley scattering
is not considered in the present work and hence we drop any reference
to the valley index. The BTE for a uniform graphene system reads as
\cite{Ziman} 
\begin{eqnarray}
\frac{\partial n_{\sigma}(\mathbf{k})}{\partial t}+\dot{\mathbf{k}}\cdot\nabla_{\mathbf{k}}n_{\sigma}(\mathbf{k}) & =\;\mathcal{I}[n_{\sigma}(\mathbf{k})]\,.\label{eq:be1}
\end{eqnarray}
In the above, $n_{\sigma}(\mathbf{k})$ is the carrier distribution
function for carriers with momentum $\mathbf{k}$ and spin projection
$\sigma$ along some axis and $\mathcal{I}[.]$ denotes the collision
integral (see below). Under an external electric field $\boldsymbol{\boldsymbol{\mathcal{E}}}$,
the BTE for carriers in the conduction (valence) band $\lambda=1$
($\lambda=-1$) becomes 
\begin{equation}
-e\lambda\boldsymbol{\boldsymbol{\mathcal{E}}}\cdot\mathbf{v}_{\mathbf{k}}^{(\lambda)}\left(\frac{\partial n^{0}}{\partial\epsilon}\right)_{\epsilon=\epsilon(\mathbf{k})}=\;\mathcal{I}[n_{\sigma}(\mathbf{k})]\,,\label{eq:first_order_Be}
\end{equation}
in first order in $\boldsymbol{\boldsymbol{\mathcal{E}}}$. Here,
$\mathbf{v}_{\mathbf{k}}^{(\lambda)}=\lambda v_{F}(\cos\theta_{\mathbf{k}},\sin\theta_{\mathbf{k}}$)
is the band velocity, $n^{0}=n^{0}(\epsilon)$ is the Fermi distribution
function evaluated at energy $\epsilon$ and $-e<0$ is the electron
charge. For simplicity we drop the band index in what follows. The
collision integral for non-interacting charge carriers reads as 
\begin{equation}
\mathcal{I}[n_{\sigma}(\mathbf{k})]=\sum_{\sigma^{\prime}=\sigma,\bar{\sigma}}\sum_{\mathbf{k}^{\prime}}\left[n_{\sigma^{\prime}}(\mathbf{k}^{\prime})-n_{\sigma}(\mathbf{k})\right]W_{\sigma^{\prime}\sigma}(\mathbf{k}^{\prime},\mathbf{k})\,,\label{eq:scattering_operator}
\end{equation}
where $W_{\sigma^{\prime}\sigma}(\mathbf{k}^{\prime},\mathbf{k})$
is the quantum-mechanical scattering probability for a process with
$\mathbf{k}^{\prime}\rightarrow\mathbf{k}$ and $\sigma^{\prime}\rightarrow\sigma$
(here, $\bar{\sigma}\equiv-\sigma$). Note that under the stated conditions
this quantity is the same in both valleys of graphene. For isotropic
Fermi surfaces the distribution function solving Eq.~(\ref{eq:first_order_Be})
has the general form 
\begin{equation}
n_{\sigma}(\mathbf{k})=n^{0}(k)+A_{\sigma}(k)|\mathbf{v}_{\mathbf{k}}|\cos[\phi(\mathbf{k})]+B_{\sigma}(k)|\mathbf{v}_{\mathbf{k}}|\sin[\phi(\mathbf{k})]\,,\label{eq:ansatz-2}
\end{equation}
where $k=|\mathbf{k}|$ and $\phi(\mathbf{x})$ denotes the angle
that $\mathbf{v}_{\mathbf{x}}$ forms with the direction of $\boldsymbol{\boldsymbol{\mathcal{E}}}$.
The functions $A_{\sigma}(k)$ and $B_{\sigma}(k)$ contain the information
needed for the calculation of steady-state (spin-dependent) currents.
Substitution of Eq.~(\ref{eq:ansatz-2}) into Eq.~(\ref{eq:first_order_Be})
yields the following system of equations 
\begin{eqnarray}
X(k) & = & \sum_{\sigma^{\prime}=\sigma,\bar{\sigma}}\left[A_{\sigma^{\prime}}\Gamma_{\sigma^{\prime}\sigma}^{\textrm{C}}-B_{\sigma^{\prime}}\Gamma_{\sigma^{\prime}\sigma}^{\textrm{S}}-A_{\sigma}\Gamma_{\sigma^{\prime}\sigma}^{\textrm{I}}\right]\,,\label{eq:f.e.1}\\
0 & = & \sum_{\sigma^{\prime}=\sigma,\bar{\sigma}}\left[B_{\sigma^{\prime}}\Gamma_{\sigma^{\prime}\sigma}^{\textrm{C}}+A_{\sigma^{\prime}}\Gamma_{\sigma^{\prime}\sigma}^{\textrm{S}}-B_{\sigma}\Gamma_{\sigma^{\prime}\sigma}^{\textrm{I}}\right]\,,\label{eq:f.e.2}
\end{eqnarray}
where $X(k)\equiv-e|\boldsymbol{\mathcal{E}}|\partial n^{0}/\partial\epsilon_{\mathbf{k}}$
and we have defined the relaxation rates: 
\begin{eqnarray}
\Gamma_{\sigma^{\prime}\sigma}^{\textrm{C}} & = & \int\frac{Sd^{2}\mathbf{k}^{\prime}}{(2\pi)^{2}}\cos\left[\phi(\mathbf{k})-\phi(\mathbf{k}^{\prime})\right]W_{\sigma^{\prime}\sigma}(\mathbf{k}^{\prime},\mathbf{k})\,,\label{eq:tau_paralell_sa-1}\\
\Gamma_{\sigma^{\prime}\sigma}^{\textrm{S}} & = & \int\frac{Sd^{2}\mathbf{k}^{\prime}}{(2\pi)^{2}}\sin\left[\phi(\mathbf{k})-\phi(\mathbf{k}^{\prime})\right]W_{\sigma^{\prime}\sigma}(\mathbf{k}^{\prime},\mathbf{k})\,,\label{eq:tau_perp_sa-1}\\
\Gamma_{\sigma^{\prime}\sigma}^{\textrm{I}} & = & \int\frac{Sd^{2}\mathbf{k}^{\prime}}{(2\pi)^{2}}W_{\sigma^{\prime}\sigma}(\mathbf{k}^{\prime},\mathbf{k})\,,\label{eq:tau_iden}
\end{eqnarray}
where $S$ denotes the area of the system. For time-reversal invariant
scattering, the relaxation rates obey $\Gamma_{\alpha\beta}^{X}=\varsigma\Gamma_{\bar{\alpha}\bar{\beta}}^{X}$
where $\varsigma=1$ ($\varsigma=-1$) for $X=I,C$ ($X=S$) \cite{symmetry}.
Using these relations, the solutions of (\ref{eq:f.e.1})-(\ref{eq:f.e.2})
can be shown to acquire a particularly simple form in terms of four
relaxation times: 
\begin{eqnarray}
A_{\sigma} & = & +A_{\bar{\sigma}}=-\frac{\tau_{\parallel}\tau_{\perp}\tau_{\perp}^{*}}{\tau_{\parallel}\tau_{\parallel}^{*}+\tau_{\perp}\tau_{\perp}}X\,,\label{eq:sol_A}\\
B_{\sigma} & = & -B_{\bar{\sigma}}=-\frac{\tau_{\perp}\tau_{\parallel}\tau_{\parallel}^{*}}{\tau_{\parallel}\tau_{\parallel}^{*}+\tau_{\perp}\tau_{\perp}}X\,,\label{eq:sol_B}
\end{eqnarray}
where 
\begin{align}
\frac{1}{\tau_{\parallel}} & =\Gamma_{\sigma\sigma}^{\textrm{I}}-\Gamma_{\sigma\sigma}^{\textrm{C}}+\Gamma_{\sigma\bar{\sigma}}^{\textrm{I}}-\Gamma_{\sigma\bar{\sigma}}^{\textrm{C}}\,,\label{eq:transp_cross}\\
\frac{1}{\tau_{\parallel}^{*}} & =\Gamma_{\sigma\sigma}^{\textrm{I}}-\Gamma_{\sigma\sigma}^{\textrm{C}}+\Gamma_{\sigma\bar{\sigma}}^{\textrm{I}}+\Gamma_{\sigma\bar{\sigma}}^{\textrm{C}}\,,\label{eq:transp_cross_star}\\
\frac{1}{\tau_{\perp}} & =\Gamma_{\sigma\sigma}^{\textrm{S}}+\Gamma_{\sigma\bar{\sigma}}^{\textrm{S}}\,,\quad\frac{1}{\tau_{\perp}^{*}}=\Gamma_{\sigma\sigma}^{\textrm{S}}-\Gamma_{\sigma\bar{\sigma}}^{\textrm{S}}\,.\label{eq:skew_cross}
\end{align}
Here, $ $$\tau_{\parallel}$ and $\tau_{\perp}$ are the standard
transport and ``skew'' relaxation times \cite{Schliemann}, whereas
$\tau_{\parallel}^{*}$ and $\tau_{\perp}^{*}$ arise due to spin
flips. Our study shows that a hierarchy of (non-equivalent) relaxation
rates emerges when a quantum number such as spin is not conserved
\cite{Lopes}. This fact has been overlooked in previous approximate
treatments of the BTE in similar systems \cite{Schliemann}. Below
we show that ``star'' relaxation times play a crucial role in the
spin Hall effect. 

For a driving electric field along the $x$ axis, the charge and spin
Hall currents are defined as 
\begin{align}
j_{x}=-eg_{v}\int\frac{Sd^{2}\mathbf{k}^{\prime}}{(2\pi)^{2}}\left[n_{\sigma}(\mathbf{k})+n_{\bar{\sigma}}(\mathbf{k})\right]v_{\mathbf{k}}\cdot\mathbf{e}_{x}\,,\label{eq:sol_JX}\\
j_{\textrm{sH}}=-eg_{v}\int\frac{Sd^{2}\mathbf{k}^{\prime}}{(2\pi)^{2}}\left[n_{\sigma}(\mathbf{k})-n_{\bar{\sigma}}(\mathbf{k})\right]v_{\boldsymbol{k}}\cdot\mathbf{e}_{y}\,,\label{eq:sol_JsH}
\end{align}
respectively, where $g_{v}=2$ is the valley degeneracy factor. At
zero temperature $X\rightarrow-e|\boldsymbol{\mathcal{E}}|\delta(\epsilon_{\mathbf{k}}-\epsilon_{F})$
the integrals over $\mathbf{k}^{\prime}$ pick up only the contribution
of states at the Fermi surface, and the spin Hall angle 
\begin{equation}
\theta_{\textrm{sH}}\equiv\frac{j_{\textrm{sH}}}{j_{x}}\,\label{eq:def_sH_angle}
\end{equation}
is totally determined by the star relaxation rates, i.e., 
\begin{equation}
\theta_{\textrm{sH}}=\frac{B_{\uparrow}-B_{\downarrow}}{A_{\uparrow}+A_{\downarrow}}=\frac{\tau_{\parallel}^{*}}{\tau_{\perp}^{*}}\,.\label{eq:spin_Hall_angle}
\end{equation}
We note that the naive formula $\theta_{\textrm{sH}}=\tau_{\parallel}/\tau_{\perp}$
can only be correct in the absence of spin-flips, in which case $\tau_{\parallel(\perp)}^{*}=\tau_{\parallel(\perp)}$.
When written in terms of cross sections, the physical interpretation
of Eq.~(\ref{eq:spin_Hall_angle}) becomes clear. Using $W_{\alpha\beta}(\mathbf{k},\mathbf{k}^{\prime})\propto\sigma_{\alpha\beta}(\theta)\delta(\epsilon_{\mathbf{k}}-\epsilon_{\mathbf{k}^{\prime}})$,
where $\sigma_{\alpha\beta}(\theta)=|f^{\alpha\beta}(\theta)|^{2}$
is the impurity differential cross section at angle $\theta\equiv\phi(\mathbf{k}^{\prime})-\phi(\mathbf{k})$
\cite{ManyImp}, we find 
\begin{equation}
\theta_{\textrm{sH}}=\frac{\Sigma_{\perp}^{\text{*}}}{\Sigma_{\parallel}^{\text{*}}}\equiv\frac{\sum_{\sigma^{\prime}=\sigma,\bar{\sigma}}\int d\theta\:\sigma_{\sigma\sigma^{\prime}}(\theta)\,\sigma\sigma^{\prime}\sin\theta}{\sum_{\sigma^{\prime}=\sigma,\bar{\sigma}}\int d\theta\:\sigma_{\sigma\sigma^{\prime}}(\theta)\left(1+\sigma\sigma^{\prime}\cos\theta\right)}\,,\label{eq:spin_Hall_angle-1}
\end{equation}
identifying $\theta_{\textrm{sH}}$ as a properly defined ``skewness'',
i.e., ratio of a skew cross section to a transport cross section.

\section{Calculation of Exact Scattering Amplitudes\label{sec:Intrinsic-scatterers}}

\subsection{Scatterers Producing Intrinsic-Type Spin-Orbit Coupling}

In this section the partial-wave scattering amplitudes $\{S_{m}\}$
for disk scatterers endowed with spin-orbit coupling (SOC) of intrinsic
type is derived. The components of the graphene spinor $\Psi^{\pm}(\mathbf{r})=(\psi_{A}^{\pm}({\mathbf{r}}),\psi_{B}^{\pm}({\mathbf{r}}))^{\textrm{T}}$
are decomposed in radial harmonics
\begin{align}
\psi_{A}^{\pm}({\mathbf{r}})= & \sum_{m=-\infty}^{\infty}g_{m,\pm}^{A}(r)e^{im\theta}\,,\label{eq:Amp_A}\\
\psi_{B}^{\pm}({\mathbf{r}})= & \sum_{m=-\infty}^{\infty}g_{m,\pm}^{B}(r)e^{i(m+1)\theta}\,,\label{eq:Amp_B}
\end{align}
where $\theta\equiv\textrm{arg}(k_{x}+ik_{y})$, $\mathbf{k}$ is
the wavevector, $m$ is the angular momentum quantum number, $\pm$
represents the spin projection and $A(B)$ are sublattice indices.
The asymptotic form of the radial functions $g_{m,\pm}^{A}(r)$ and
$g_{m,\pm}^{B}(r)$ determine the scattering amplitudes. The Hamiltonian
is 
\begin{equation}
\mathcal{H}=\mathcal{H}_{0}+\left(V_{0}+\Delta_{I}\tau_{z}\sigma_{z}s_{z}\right)\Theta(R-r)\,,\label{eq:Ham_Intrinsic}
\end{equation}
where $\mathcal{H}_{0}=v_{F}(\tau_{z}\sigma_{x}p_{x}+\sigma_{y}p_{y})$
is the low-energy free Hamiltonian and the second term is the disk
scatterer potential. Here, $\Theta(.)$ is the Heaviside step function
and $\boldsymbol{\mathbf{\sigma}},\mathbf{\boldsymbol{\tau}}$ and
$\mathbf{s}$ are Pauli matrices for sublattice, valley and spin,
respectively. We set $\hbar\equiv1$ througout. The asymptotic form
of waves at the $K$ valley ($\tau_{z}=1$) having spin projection
$s=\mathbf{s}\cdot\mathbf{e}_{z}$ is
\begin{align}
|\psi_{\lambda,\mathbf{k},s}({\bf r})\rangle & =\left(\begin{array}{c}
1\\
\lambda
\end{array}\right)e^{ikr\cos(\theta)}|s\rangle+\frac{f^{ss}(\theta)}{\sqrt{-ir}}\left(\begin{array}{c}
1\\
\lambda e^{i\theta_{{\bf k}}}
\end{array}\right)e^{ikr}|s\rangle\nonumber \\
 & +\frac{f^{s\bar{s}}(\theta)}{\sqrt{-ir}}\left(\begin{array}{c}
1\\
\lambda e^{i\theta_{{\bf k}}}
\end{array}\right)e^{ikr}|\bar{s}\rangle\ ,\label{eq:psiKas-1}
\end{align}
where $\lambda=\pm1$ denotes the carrier polarity, $\bar{s}=-s$
and $f^{ss}(\theta)$ and $f^{s\bar{s}}(\theta)$ are scattering amplitudes
in the elastic and spin-flip channels, respectively. (For other choices
of quantization axis see discussion in Sec.~\ref{sec:QA}.) Inside
the disk of radius $R$, the dispersion relation satisfies $E-V_{0}=\lambda\sqrt{v_{F}^{2}k^{2}+\Delta_{I}^{2}}\equiv\epsilon$
and

\begin{equation}
|\psi_{\lambda,{\bf k},s}(\mathbf{r})\rangle=\left(\begin{array}{c}
\sqrt{\epsilon+s\Delta_{I}}\\
\eta\sqrt{\epsilon-s\Delta_{I}}e^{i\theta_{{\bf k}}}
\end{array}\right)e^{i{\bf {k}\cdot{\bf {r}}}}|s\rangle,\label{psiKso}
\end{equation}
where $\eta=\mathrm{sgn}(\epsilon+|\Delta_{I}|)$. In order to identify
the scattering amplitudes, we recast the wavefunction inside and outside
the disk as a superposition of angular harmonics. For $r>R$, we have
$E=\lambda v_{F}k$ and the partial-wave $m$ is given by
\begin{align}
|\psi_{m}^{>}(r,\theta)\rangle & =\left(\begin{array}{c}
J_{m}(kr)e^{im\theta}\\
i\lambda J_{m+1}(kr)e^{i(m+1)\theta}
\end{array}\right)|s\rangle\nonumber \\
 & +S_{m}^{s}\left(\begin{array}{c}
H_{m}^{(1)}(kr)e^{im\theta}\\
i\lambda H_{m+1}^{(1)}(kr)e^{i(m+1)\theta}
\end{array}\right)|s\rangle,\label{eq:psi_outside}
\end{align}
whereas for $r<R$ one has 
\begin{equation}
|\psi_{m}^{<}(r,\theta)\rangle=C_{m}\left(\begin{array}{c}
\sqrt{\epsilon+s\Delta_{I}}J_{m}(\beta r)e^{im\theta}\\
i\eta\sqrt{\epsilon-s\Delta}J_{m+1}(\beta r)e^{i(m+1)\theta}
\end{array}\right)|s\rangle,\label{eq:psi_inside}
\end{equation}
with $\beta\equiv\sqrt{\epsilon^{2}-\Delta^{2}}/v_{F}$. The boundary
condition $\psi_{m}^{>}(R,\theta)=\psi_{m}^{<}(R,\theta)$ gives rise
to two equations fully determining the amplitudes $S_{m}^{s}$. Straighforward
algebra yields\begin{widetext}

\begin{equation}
S_{m}^{s}=-\frac{\sqrt{\epsilon+s\Delta_{I}}J_{m+1}(kR)J_{m}(\beta R)-\frac{\eta}{\lambda}\sqrt{\epsilon-s\Delta_{I}}J_{m+1}(\beta R)J_{m}(kR)}{\sqrt{\epsilon+s\Delta_{I}}H_{m+1}^{(1)}(kR)J_{m}(\beta R)-\frac{\eta}{\lambda}\sqrt{\epsilon-s\Delta_{I}}J_{m+1}(\beta R)H_{m}^{(1)}(kR)}.\label{phase-shifts-SOI}
\end{equation}
\end{widetext}

Naturally, in the absence of intervalley scattering $[\tau_{z},\mathcal{H}]=0$,
calculations performed in the $K$ and $K^{\prime}$ valleys yield
the same scattering amplitudes and hence the same transport quantities
\cite{valleys}.

\subsection{Scatterers Producing Rashba-Type Spin-Orbit Coupling}

If we consider a scatterer producing a Rashba-type SOC interaction
in the form 
\begin{equation}
\tilde{\mathcal{V}}=\left[V_{0}+\tau_{z}\Delta_{R}\left(\sigma_{x}s_{y}-\sigma_{y}s_{x}\right)\right]\Theta(R-r)\,,\label{eq:tilde_V}
\end{equation}
the diagonalization of $\tilde{\mathcal{H}}=\mathcal{\tilde{\mathcal{H}}}_{0}+\tilde{\mathcal{V}}$
inside the disk ($r<R$) yields the spectrum $E-V_{0}=\xi\tau_{z}\Delta_{R}+\lambda\sqrt{v_{F}^{2}k^{2}+\Delta_{R}^{2}}=\epsilon_{\xi}(\mathbf{k})\equiv\epsilon_{\xi}$,
where $\xi=\pm$ is the chirality of the band~\cite{Hernando09_sup}.
For simplicity we restrict the subsequent analysis to carriers with
positive polarity $\lambda=1$ and assume $|\epsilon|>2|\Delta_{R}|$.
Eigenstates at the $K$ valley read as 
\begin{equation}
|\psi_{\mathbf{k}}(\mathbf{r})\rangle=\left[\left(\begin{array}{c}
1\\
\frac{\epsilon_{\xi}}{v_{F}k}e^{i\theta_{{\bf k}}}
\end{array}\right)|\uparrow\;\rangle+i\xi\left(\begin{array}{c}
\frac{\epsilon_{\xi}}{v_{F}k}e^{i\theta_{{\bf k}}}\\
e^{2i\theta_{{\bf k}}}
\end{array}\right)|\downarrow\;\rangle\right]e^{i{\bf {k}\cdot{\bf {r}}}}.\label{psiKR}
\end{equation}
Differently from the intrinsic SOC, Rashba-like interaction entangles
spin and and pseudo-spin (sublattice), implying that spin-flips must
be taken into account. As before, eigenstates inside and outside the
disk scatterer can be recast into a superposition of angular harmonics.
In the region $r>R$ we obtain
\begin{align}
|\psi_{m}^{>}(r,\theta)\rangle & =\left(\begin{array}{c}
J_{m}(kr)e^{im\theta}\\
iJ_{m+1}(kr)e^{i(m+1)\theta}
\end{array}\right)|\uparrow\;\rangle\nonumber \\
 & +S_{m}^{\uparrow\uparrow}\left(\begin{array}{c}
H_{m}^{(1)}(kr)e^{im\theta}\\
iH_{m+1}^{(1)}(kr)e^{i(m+1)\theta}
\end{array}\right)|\uparrow\;\rangle\notag\\
 & +S_{m+1}^{\uparrow\downarrow}\left(\begin{array}{c}
H_{m+1}^{(1)}(kr)e^{i(m+1)\theta}\\
iH_{m+2}^{(1)}(kr)e^{i(m+2)\theta}
\end{array}\right)|\downarrow\;\rangle.\label{eq:rashbaout}
\end{align}
 In the above we assumed an incident wave with $s=1$. Inside the
disk, the wave function regular at the origin is
\begin{align}
|\psi_{m}^{<}(r,\theta)\rangle=\sum_{\xi}C_{\xi m} & \left[\left(\begin{array}{c}
J_{m}(\beta_{\xi}r)e^{im\theta}\\
i\frac{\epsilon_{\xi}}{v_{F}\beta_{\xi}}J_{m+1}(\beta_{\xi}r)e^{i(m+1)\theta}
\end{array}\right)|\uparrow\;\rangle\right.\nonumber \\
 & \left.+\xi\left(\begin{array}{c}
\frac{\epsilon}{v_{F}\beta_{\xi}}J_{m+1}(\beta_{\xi}r)e^{i(m+1)\theta}\\
iJ_{m+2}(\beta_{\xi}r)e^{i(m+2)\theta}
\end{array}\right)|\downarrow\;\rangle\right]\,,\label{eq:rashbainside}
\end{align}
where $\beta_{\xi}=\sqrt{\epsilon_{\xi}(\epsilon_{\xi}-2\xi\Delta_{R})}/v_{F}$.
The matching conditions at $r=R$ yields four equations 
\begin{align}
J_{m}(kR)+S_{m}^{\uparrow\uparrow}H_{m}^{(1)}(kR) & =\sum_{\xi}C_{\xi m}J_{m}(\beta_{\xi}R),\label{eq:M1}\\
J_{m}(kR)+S_{m}^{\uparrow\uparrow}H_{m}^{(1)}(kR) & =\sum_{\xi}C_{\xi m}J_{m}(\beta_{\xi}R),\label{eq:M2}\\
S_{m}^{\uparrow\downarrow}H^{(1)}(kR) & =\sum_{\xi}\xi\frac{\epsilon C_{\xi m}}{v_{F}\beta_{\xi}}J_{m+1}(\beta_{\xi}R),\label{eq:M3}\\
S_{m}^{\uparrow\downarrow}H_{m+2}^{(1)}(kR) & =\sum_{\xi}\xi C_{\xi m}J_{m+2}(\beta_{\xi}R).\label{eq:M4}
\end{align}
Equations~(\ref{eq:M1})--(\ref{eq:M4}) can be shown to obey the
required boundary conditions. Indeed, taking a superposition of partial
waves $\psi=\sum_{m}i^{m}\psi_{m}$ the correct asymptotic limit for
the Dirac equation in two dimensions is obtained, i.e.,
\begin{align}
|\psi_{\mathbf{k}}(\mathbf{r})\rangle\rightarrow & \left(\begin{array}{c}
1\\
1
\end{array}\right)e^{ikr\cos\theta}|\uparrow\;\rangle+\sqrt{\frac{2}{i\pi kr}}\left(\begin{array}{c}
1\\
e^{i\theta}
\end{array}\right)e^{ikr}\times\nonumber \\
 & \sum_{m=-\infty}^{\infty}e^{im\theta}\left[{S_{m}^{ss}}|\uparrow\;\rangle+{S_{m}^{s\bar{s}}}|\downarrow\;\rangle\right]\,.\label{eq:asympt}
\end{align}
The scattering amplitudes can be readily identified from the above
expression: 
\begin{eqnarray}
f^{ss}(\theta) & = & \sqrt{\frac{2}{i\pi k}}\sum_{m=-\infty}^{\infty}{S_{m}^{ss}e^{im\theta}},\label{eq:f_elastic}\\
f^{s\bar{s}}(\theta) & = & \sqrt{\frac{2}{i\pi k}}\sum_{m=-\infty}^{\infty}{S_{m}^{s\bar{s}}e^{im\theta}}.\label{eq:f_spin_flip}
\end{eqnarray}

\subsection{General Expressions of Cross Sections }

The formulae given above allows determination of cross sections (or
equivalently, relaxation rates) used in the BTE (Sec.~\ref{sec:BTE}).
For instance, the ``star'' transport and the ``star'' skew cross
sections 
\begin{align}
\Sigma_{\parallel}^{*} & =\sum_{s^{\prime}}\int d\theta(1-ss^{\prime}\cos\theta)|f^{ss^{\prime}}(\theta)|^{2}\,,\label{eq:sigma_transp_def}\\
\Sigma_{\perp}^{*} & =\sum_{s^{\prime}}\int d\theta\sin\theta ss^{\prime}|f^{ss^{\prime}}(\theta)|^{2}\,,\label{eq:sigma_skew_def}
\end{align}
are conveniently written in terms of scattering amplitudes as 
\begin{eqnarray}
\Sigma_{\parallel}^{*} & = & \frac{4}{k}\sum_{s^{\prime}}\sum_{m=-\infty}^{\infty}\left\{ |S_{m}^{ss^{\prime}}|^{2}-ss^{\prime}{\rm {Re}}[S_{m}^{ss^{\prime}}(S_{m+1}^{ss^{\prime}})^{*}]\right\} \,,\label{eq:sigma_transp}\\
\Sigma_{\perp}^{*} & = & \frac{4}{k}\sum_{s^{\prime}}\sum_{m=-\infty}^{\infty}ss^{\prime}{\rm {Im}}[S_{m}^{ss^{\prime}}(S_{m+1}^{ss^{\prime}})^{*}].\label{eq:sigma_skew}
\end{eqnarray}
These expressions together with the equations defining the scattering
amplitudes explicitly {[}e.g., Eq.~(\ref{phase-shifts-SOI}){]} were
used to create the plots of the skewness and spin Hall angle shown
in the main text of the Letter.

\section{Additional Discussions\label{sec:Discussions}}

\subsection{Time-Reversal Symmetry Breaking}

In order to assess how time-reversal symmetry breaking potentially
impacts on the spin Hall effect, it is enough to add a local exchange
field $\mathcal{H}_{B}=\Delta_{B}s_{z}\Theta(R-r)$ to Eq.~(\ref{eq:Ham_Intrinsic})
and compute the spin Hall angle. Using the representation $\langle\mathbf{r}|\tilde{\Psi}\rangle$
referred to in Ref.~\onlinecite{valleys}, we obtain 
\begin{align}
\tilde{\mathcal{H}} & =\tilde{\mathcal{H}}_{0}+\left(V_{0}+\Delta_{I}\sigma_{z}s_{z}+\Delta_{B}s_{z}\right)\Theta(R-r)\,.\label{eq:Ham_BrokenTR}
\end{align}
The dispersion relation for $r<R$ satisfies $E-V_{0}-s\Delta_{B}=\tau_{z}\lambda\sqrt{\Delta_{I}^{2}+v_{F}^{2}k^{2}}\equiv\epsilon_{s}$.
As in above, for simplicity we particularize our discussion to the
conduction band $\lambda=1$. The eigenstates inside the disk read
as . 
\begin{equation}
|\psi_{m}^{<}(r,\theta)\rangle=C_{m}\left(\begin{array}{c}
\sqrt{\epsilon_{s}+s\Delta_{I}}J_{m}(\beta_{s}r)e^{im\theta}\\
i\tau\zeta_{\tau}\sqrt{\epsilon_{s}-s\Delta_{I}}J_{m+1}(\beta_{s}r)e^{i(m+1)\theta}
\end{array}\right)|s\rangle\,,\label{eq:psi_in_exchange}
\end{equation}
with $\tau\equiv\tau_{z}$, $\zeta_{\tau}=\textrm{sign}(\epsilon_{s}-|\Delta_{I}|)$
and $\beta_{s}=\sqrt{\epsilon_{s}^{2}-\Delta_{I}^{2}}/v_{F}$ Outside
the disk we find
\begin{align}
|\psi_{m}^{>}(r,\theta)\rangle & =\left(\begin{array}{c}
J_{m}(kr)e^{im\theta}\\
i\tau J_{m+1}(kr)e^{i(m+1)\theta}
\end{array}\right)|s\rangle\nonumber \\
 & +S_{m}^{s\tau}\left(\begin{array}{c}
H_{m}^{(1)}(kr)e^{im\theta}\\
i\tau H_{m+1}^{(1)}(kr)e^{i(m+1)\theta}
\end{array}\right)|s\rangle.\label{eq:psi_out_exchange-1}
\end{align}
 The skewness (or equivalently, the spin Hall angle at zero temperature)
is given by 
\begin{equation}
\gamma=\frac{B_{\uparrow}-B_{\downarrow}}{A_{\uparrow}+A_{\downarrow}}\,,\label{eq:skewness_BreakingTR}
\end{equation}
where $A_{\uparrow}\neq A_{\downarrow}$ and $B_{\uparrow}\neq-B_{\downarrow}$
for $\Delta_{B}\neq0$. In the cases of interest the smallest energy
scale will be the SOC (in the range 1--10~meV; see main text). We
have verified that near resonances large $\gamma$ is obtained even
in the strong exchange field limit $|\Delta_{B}|\gg|\Delta_{I}|$.
This simple calculation illustrates that skew scattering is robust
with respect to time-reversal symmetry breaking e.g., via local magnetic
moments sitting at the SOC-active impurity sites.

\subsection{Interference Between Intrinsic and Rashba-Type Spin-Orbit Couplings}

We now briefly discuss the robustness of the spin Hall effect with
respect to admixture of SOC terms. In realistic scenarios adsorbed
species in graphene will give rise to local SOC terms with different
symmetries, such as intrinsic and Rashba-type SOC. 

\begin{figure}[h]
\begin{centering}
\includegraphics[clip,width=0.95\columnwidth]{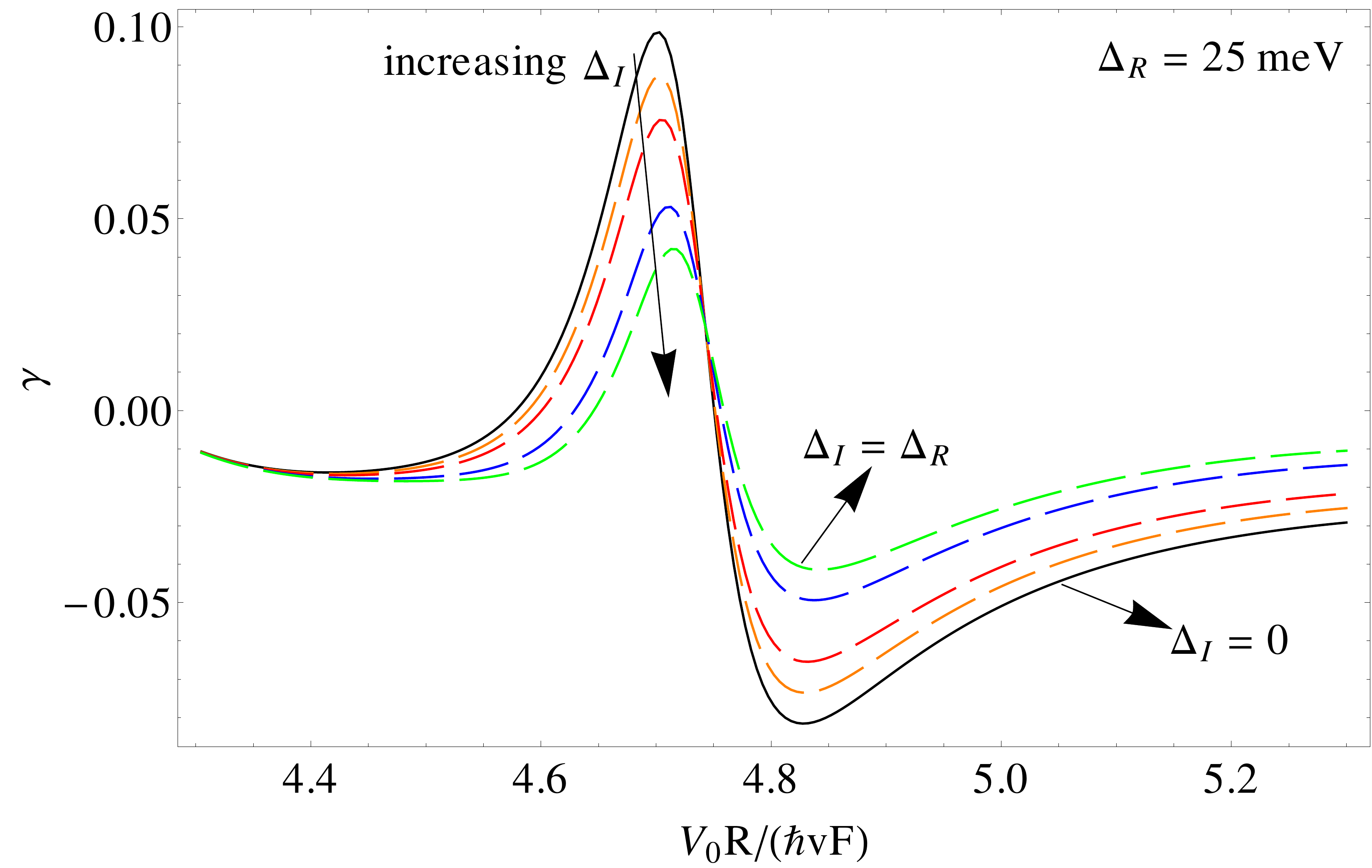}
\par\end{centering}

\begin{centering}
\includegraphics[clip,width=0.95\columnwidth]{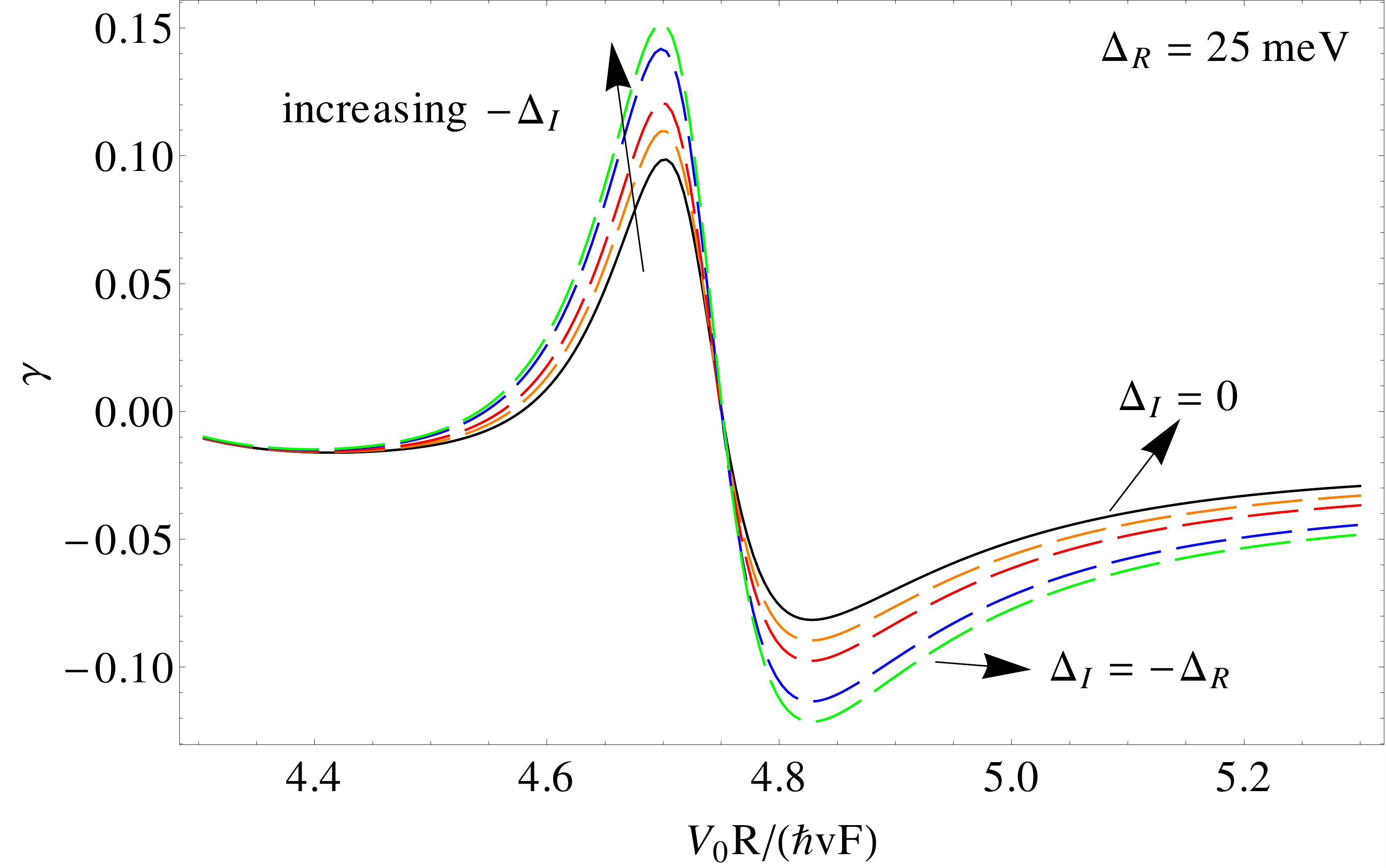} 
\par\end{centering}

\caption{\label{fig:fig1-sup}The skewness as function of the normalized eletrostatic
potential for a disk scatterer producing an admixture of intrinsic
and Rashba SOC. Values of intrinsic-type SOC are $\pm\{5,10,15,20,25\}$~meV
{[}positive (negative) values are shown in left (right) panels{]}.
Other parameters as in Fig.~2 in the manuscript.}
\end{figure}
We consider the following model: 
\begin{equation}
\tilde{\mathcal{H}}=\tilde{\mathcal{H}}_{0}+\left[V_{0}+\Delta_{I}\sigma_{z}s_{z}+\Delta_{R}\tau_{z}(\sigma_{y}s_{x}-\sigma_{x}s_{y})\right]\Theta(R-r)\,.\label{eq:hydrid_SOC_scatterer}
\end{equation}
Diagonalization inside the disk of radius $R$ yields 
\begin{equation}
E-V_{0}+\xi\Delta_{R}+\lambda\sqrt{v_{F}^{2}k^{2}+\left(\Delta_{I}-\xi\Delta_{R}\right)^{2}}\equiv\epsilon_{\chi}(\mathbf{k})\equiv\epsilon\,.\label{eq:en_chi_s_k_Rashba.V.eq.0-1}
\end{equation}
The (non-normalized) eigenvectors in the $K$ valley (and for $\lambda=1$)
can be written as 
\begin{equation}
|\psi_{\xi}(\mathbf{r})\rangle=\left[\left(\begin{array}{c}
e^{-i\theta_{\mathbf{k}}}\\
\frac{\epsilon_{\chi}(k)-\Delta_{I}}{v_{F}k}
\end{array}\right)|\uparrow\;\rangle+i\xi\left(\begin{array}{c}
\frac{\epsilon_{\chi}(k)-\Delta_{I}}{v_{F}k}\\
e^{i\theta_{\mathbf{k}}}
\end{array}\right)|\downarrow\;\rangle\right]e^{i\mathbf{k}\cdot\mathbf{r}}\,.\label{eq:eigen_inside_disk-1}
\end{equation}
Following the same procedure as outlined in the previous sections,
we find the following set of equations: 
\begin{align}
J_{m}(kR)+S_{m}^{\uparrow\uparrow}H_{m}^{(1)}(kR) & =\sum_{\xi}C_{\xi m}J_{m}(\beta_{\xi}R),\label{eq:M1-1}\\
J_{m}(kR)+S_{m}^{\uparrow\uparrow}H_{m}^{(1)}(kR) & =\sum_{\xi}C_{\xi m}J_{m}(\beta_{\xi}R),\label{eq:M2-1}\\
S_{m}^{\uparrow\downarrow}H^{(1)}(kR) & =\sum_{\xi}\xi\frac{\epsilon-\Delta_{I}}{v_{F}\beta_{\xi}}C_{\xi m}J_{m+1}(\beta_{\xi}R),\label{eq:M3-1}\\
S_{m}^{\uparrow\downarrow}H_{m+2}^{(1)}(kR) & =\sum_{\xi}\xi C_{\xi m}J_{m+2}(\beta_{\xi}R),\label{eq:M4-1}
\end{align}
where $\beta_{\xi}=\sqrt{(\epsilon-\xi\Delta_{R})^{2}-(\Delta_{I}-\xi\Delta_{R})^{2}}$.
The competition of intrinsic and Rashba couplings in the vicinity
of a resonance is demonstrated in Fig.~\eqref{fig:fig1-sup}. We
found that in general interference between SOC couplings do not supress
the resonant enhancement of the skewness.

\subsection{Quantization Axis: Arbitrary Direction of the Spin Polarization\label{sec:QA}}

In the main text of the Letter we have chosen to present our results
with spin quantization axis along the $z$ direction. However, they
can be easily generalized to any quantization direction. Physically,
as we are dealing with unpolarized currents in the spin Hall effect,
an arbitrary change in the quantization axis correspond to a measurement
of the spin polarization in an arbitrary direction. The spin-dependent
scattering amplitudes can be recast into matrix form: 
\begin{equation}
\mathbf{F}=\left(\begin{array}{cc}
f^{\uparrow\uparrow}(\theta) & f^{\uparrow\downarrow}(\theta)\\
f^{\downarrow\uparrow}(\theta) & f^{\downarrow\downarrow}(\theta)
\end{array}\right).
\end{equation}
Changing the spin quantization axis translates into a rotation in
the spin space $\mathbf{F}^{\prime}=\mathbf{U}^{-1}\mathbf{F}\mathbf{U}.$
As an example, let us consider the calculation of the spin polarization
in the $x$ direction. In this case, $\mathbf{U}$ is a $2\times2$
Hadamard matrix, i.e., 
\begin{equation}
\mathbf{H}=\mathbf{H}^{-1}=\frac{1}{\sqrt{2}}\left(\begin{array}{cc}
1 & 1\\
1 & -1
\end{array}\right).
\end{equation}
After performing the rotation, we find\begin{widetext} 
\begin{equation}
\mathbf{F}_{x}=\frac{1}{2}\left(\begin{array}{cc}
f^{\uparrow\uparrow}(\theta)+f^{\uparrow\downarrow}(\theta)+f^{\downarrow\uparrow}(\theta)+f^{\downarrow\downarrow}(\theta) & f^{\uparrow\uparrow}(\theta)+f^{\uparrow\downarrow}(\theta)-f^{\downarrow\uparrow}(\theta)-f^{\downarrow\downarrow}(\theta)\\
f^{\uparrow\uparrow}(\theta)-f^{\uparrow\downarrow}(\theta)+f^{\downarrow\uparrow}(\theta)-f^{\downarrow\downarrow}(\theta) & f^{\uparrow\uparrow}(\theta)-f^{\uparrow\downarrow}(\theta)-f^{\downarrow\uparrow}(\theta)+f^{\downarrow\downarrow}(\theta)
\end{array}\right).
\end{equation}
\end{widetext}Moreover, using 
\begin{eqnarray}
f_{x}^{ss}(\theta) & = & \sqrt{\frac{2}{i\pi k}}\sum_{m=-\infty}^{\infty}{S_{m,x}^{ss}e^{im\theta}},\label{eq:}\\
f_{x}^{s\bar{s}}(\theta) & = & \sqrt{\frac{2}{i\pi k}}\sum_{m=-\infty}^{\infty}{S_{m,x}^{s\bar{s}}e^{im\theta}},
\end{eqnarray}
the new amplitudes $S_{m,x}^{ss^{\prime}}$ can be written in terms
of the amplitudes that were calculated in the previous sections. As
a result, the ``star'' cross sections in the new quantization axis
can be obtained by using the relations given by equations \ref{eq:sigma_transp}
and \ref{eq:sigma_skew}. Our calculations show that $\Delta_{I}$-scatterers
give rise to zero skewness for carriers spin-polarized along $x$.
On the other hand, $\Delta_{R}$-scatterers produces skew-scattering
cross sections of the same order of magnitude than those for carriers
spin-polarized along $z$. Physically, it means is that in order to
measure the spin Hall effect produced by intrinsic-type SOC, it is
necessary to detect the spin-polarization in the $z$ direction while
a measurement of the spin-polarization in $x$ only detects the spin
Hall effect due to Rashba.

\section{Limitations of Perturbative Approaches: The Distorted-Wave Born Approximation}

\begin{figure}[h]
\centering{}\includegraphics[width=1\columnwidth]{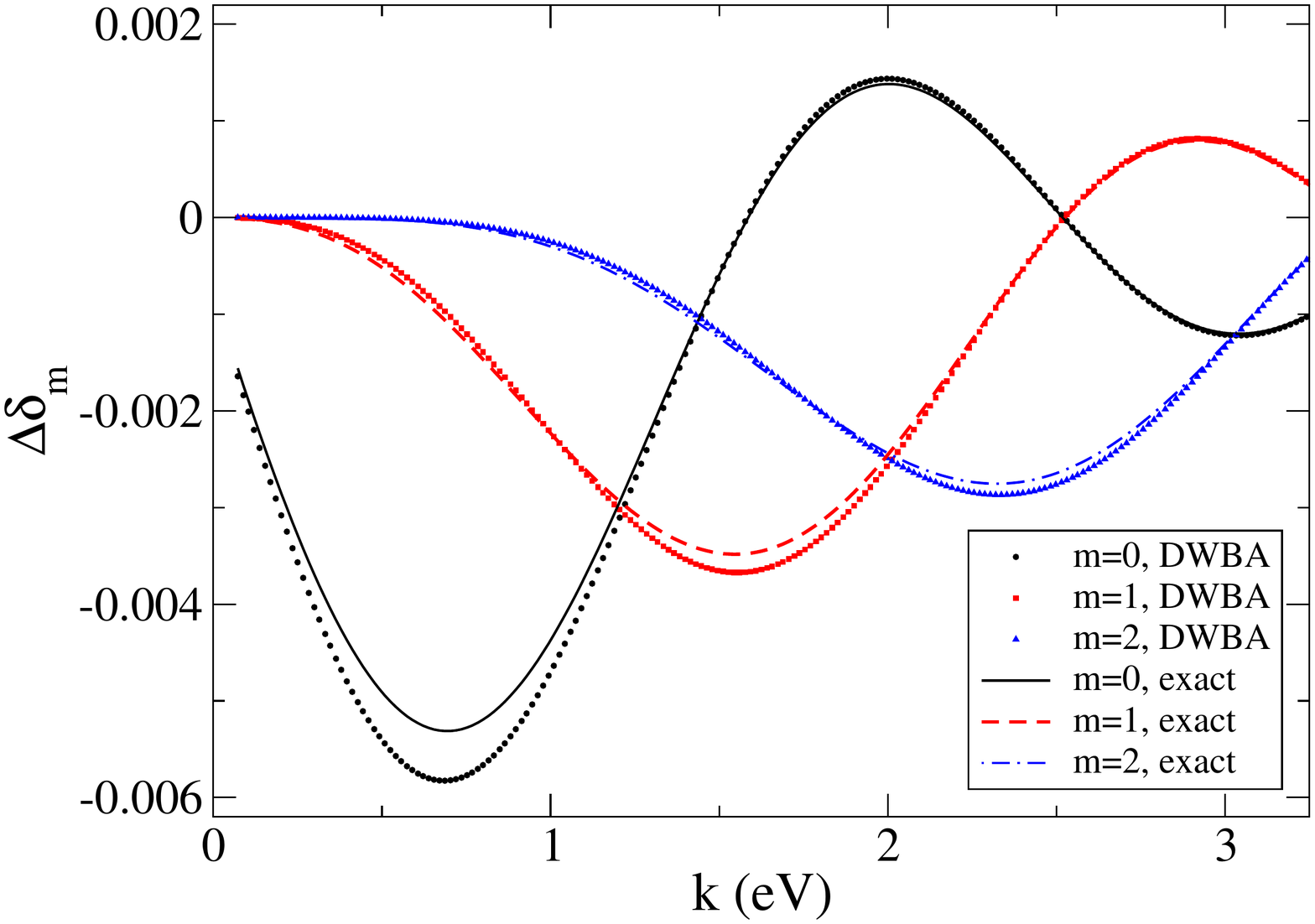} \caption{\label{fig:fig2-sup} Comparison between the exact value of $\Delta\delta_{m}=\delta_{V_{0}+\textrm{SO}}-\delta_{V_{0}}$
and the DWBA result. The parameters used in this plot are: $\Delta_{I}s_{z}=7$~eV,
$V_{0}=70$~meV and $R=1$~nm.}
\end{figure}

In this section, we provide a few examples of the limitations of the
distorted-wave Born approximation (DWBA) when applied to study spin
Hall effect (SHE) in graphene. We first derive the DWBA for a general
class of potentials of the form 
\begin{equation}
\mathcal{V}_{a}=V_{a}(r)+W_{a}(r)\sigma_{z}\,,\label{eq:potential_DWBA}
\end{equation}
where $W_{a}(r)$ denotes the sublattice symmetry breaking term. In
analogy to the derivation for a scalar potential in the Schrödinger
equation~\cite{Ballantine_sup}, it is necessary to write two copies
of the Dirac equation corresponding to different potentials, $V_{1}$
and $V_{2}$. The scattering amplitudes (or phase-shifts, $\delta_{m}^{[1]}$)
of the simpler problem $\mathcal{H}_{1}\equiv\mathcal{H}_{0}+\mathcal{V}_{1}$
are assumed to be known. Let us denote the eigenstates of $\mathcal{H}_{a}\equiv\mathcal{H}_{0}+V_{a}$
in a given valley by 
\begin{equation}
\Psi_{m}^{a}(r,\phi)=e^{im\phi}\left(\begin{array}{c}
F_{a}(r)\\
G_{a}(r)e^{i\phi}
\end{array}\right),\label{eq:spinor_ansatz}
\end{equation}
and $a=1,2$. We aim at finding the phase-shifts induced by the sublattice
breaking term $W(r)$. Inserting the ansatz (\ref{eq:spinor_ansatz})
into the Dirac equation, and using the asymptotic form of graphene
wavefunctions 
\begin{eqnarray}
\Psi_{m}^{a}(r,\phi) & \rightarrow & \sqrt{\frac{2}{\pi kr}}\left[\begin{array}{c}
\cos\left(kr-\varphi_{m}+\delta_{m}^{a}\right)\\
\lambda i\sin\left(kr-\varphi_{m}+\delta_{m}^{a}\right)
\end{array}\right]\,,\label{eq:asymp_F}
\end{eqnarray}
where $\varphi_{m}=(2m+1)\pi/4$, we find
\begin{align}
\frac{2\lambda}{\pi k}\sin\left(\delta_{m}^{[1]}-\delta_{m}^{[2]}\right) & =\int_{0}^{\infty}drr\left[\frac{\delta\mathcal{V}_{+}}{\hbar v_{F}}F_{1}(r)F_{2}(r)\right.\nonumber \\
 & \left.-\frac{\delta\mathcal{V}_{-}}{\hbar v_{F}}G_{1}(r)G_{2}(r)\right]\,,\label{eq:aux_DWBA}
\end{align}
where $\delta\mathcal{V}_{\pm}\equiv V_{2}\pm W_{2}-(V_{1}\pm W_{1})$.
The above result is still exact; the DWBA is derived by employing
the ``Born approximation'': $F_{2}(r)\simeq F_{1}(r)$ and $G_{2}(r)\simeq G_{1}(r)$.
Specializing to the case of interest, i.e.,~$V_{1}=V_{2}=V_{0}(r)$,
$W_{1}=0$, and $W_{2}=W(r)$, the DWBA yields 
\begin{equation}
\Delta\delta_{m}=-\frac{\zeta\pi k}{2\hbar v_{F}}\int_{0}^{\infty}drrW(r)\left[f_{m}(r)^{2}+g_{m}(r)^{2}\right],\label{eq:DWBA}
\end{equation}
where $\Delta\delta_{m}=\delta_{m}^{[2]}-\delta_{m}^{[1]}$ is the
correction to the $m$-th phase-shift $\delta_{m}^{[1]}$ introduced
by the sublattice breaking term $W(r)$. In the above, $f_{m}(g_{m})$
are the partial-wave amplitudes of the simpler problem $\mathcal{H}_{1}=\mathcal{H}_{0}+V_{0}(r)$
and $\zeta=\mathrm{sgn}(E-V_{0})$. We can use the equation above
to calculate $\Delta\delta_{m}$ explicitly for an intrinsic-type
disk scatterer with $V_{0}(r)=V_{0}\Theta(R-r)$ and $W(r)=\Delta_{I}s_{z}\Theta(R-r)$.
We find 
\begin{equation}
\Delta\delta_{m}=-\frac{\zeta\pi s_{z}}{2}\frac{\Delta_{I}k}{\hbar v_{F}\alpha^{2}}\int_{0}^{\alpha R}duu\left[J_{m}(u)^{2}-J_{m+1}(u)^{2}\right].
\end{equation}
where $\alpha\equiv|k-V_{0}/{\hbar v_{F}}|$. The above expression
can be further simplified using the properties of Bessel functions
(not shown). In Fig.~\ref{fig:fig2-sup} we can see the comparison
between this approximation and the exact result using Eq.~\eqref{phase-shifts-SOI}
and the relation $S_{m}^{s}=ie^{i\delta_{m}^{s}}\sin(\delta_{m}^{s})$.
The skew cross section can be easily calculated under the DWBA: 
\begin{equation}
\Sigma_{\perp}^{\textrm{DWBA}}=\frac{2}{k}\sum_{m=-\infty}^{\infty}(\Delta\delta_{m}-\Delta\delta_{m+1})\cos[2(\delta_{m}^{[1]}-\delta_{m+1}^{[1]})].
\end{equation}

The DWBA seems promising to compute phase-shifts for intrinsic-type
scatterers in the presence of a scalar potential. However, it fails
to correctly describe the skew cross section (and thus SHE) for other
symmetries or in the presence of resonant scattering. Here, we briefly
discuss a few situations where the approximation is not valid. Our
first example is provided by a void in graphene, which is described
by the boundary condition requiring that the A-sublattice component
of the spinor $|\psi_{\mathbf{k}}(\mathbf{r})\rangle$ vanishes at
$r=R$. Hence, the spin-independent part of the scattering phase shift
$\delta_{m}^{[1]}$ fullfils: 
\begin{equation}
\tan\delta_{m}^{[1]}=\frac{J_{m}(kR)}{Y_{m}(kR)}.
\end{equation}
Note the symmetry $\delta_{-m}^{[1]}=\delta_{m}^{[1]}$. If we assume
that the intrinsic-type potential only acts in the edge of the void,
i.e.,~$W(r)=R\Delta_{I}\delta(r-R)\tau_{z}\sigma_{z}s_{z}$, then,
the DWBA gives 
\begin{align}
\Delta\delta_{m} & \propto\int rdr\:\Psi_{m}^{\dag}(r,\phi)W(r)\Psi_{m}(r,\phi)\notag\\
 & =R\Delta_{I}\frac{\left[Y_{m}(kR)J_{m+1}(kR)-Y_{m+1}(kR)J_{m}(kR)\right]^{2}}{Y_{m}^{2}(kR)}\notag\\
 & \propto\Delta_{I}\,,
\end{align}
where we have used the Wronskian identity for Bessel function, which
implies that $Y_{m}(x)J_{m+1}(x)-Y_{m+1}(x)J_{m}(x)=2/(\pi x)$. Hence,
within the DWBA, $\delta_{m}^{[2]}=\delta_{-m}^{[2]}$, that is, the
same symmetry as for the void potential, which implies the absence
of skew scattering and therefore SHE.

A second example is provided by a generic Rashba-type scatterer, for
which $W(r)=\Delta_{R}(r)\left(\tau^{z}\sigma^{x}s^{y}-\sigma^{y}s^{x}\right)$.
It can be shown that within the DWBA, and at the lowest order in $\Delta_{R}$,
only the spin-flip amplitude $f^{s\bar{s}}(\theta)\neq0$ gets corrected
and therefore the skew cross section for $z$-polarization (and hence
SHE) is zero in this approximation, just as in the previous example.


\begin{thebibliography}{10}
\bibitem{BTE_comment}We have neglected spin coherence in $n_{s}(\mathbf{k})$,
keeping only diagonal terms, which is expected to be a good approximation
at room temperature. We further note that the BTE does not describe
transport close to the Dirac point.
\end{thebibliography}
\begin{thebibliography}{10}
\bibitem{symmetry}These relations can be easily shown invoking $W_{\sigma\sigma^{\prime}}(\mathbf{k},\mathbf{k}^{\prime})=|T_{\sigma^{\prime}\sigma}(\mathbf{k}^{\prime},\mathbf{k})|^{2}$
and exploiting the symmetries of the $T$ matrix; see Eq. (3) in the
main text of the Letter and comments therein.
\bibitem{ManyImp}When multiple disorder sources are present, weighted
summations are taken in the usual way: $W_{\alpha\beta}(\mathbf{k},\mathbf{\mathbf{k}^{\prime}})=\sum_{j}W_{\alpha\beta}^{j}(\mathbf{k},\mathbf{k}^{\prime})\propto\sum_{j}n^{(j)}\sigma_{\alpha\beta}^{j}(\theta)\delta(\epsilon_{\mathbf{k}}-\epsilon_{\mathbf{k}^{\prime}})$
with $n^{(j)}$ being the areal density of impurities of type $j$.
\bibitem{valleys}In fact, using the representation $|\tilde{\Psi}\rangle=(\psi_{A}^{K},\psi_{B}^{K},\psi_{B}^{K^{\prime}},\psi_{A}^{K^{\prime}})^{\textrm{T}}\otimes|s\rangle$
yields an effective $\mathcal{H}$ in the form of two copies of the
Dirac Hamiltonian $\tilde{\mathcal{H}}_{0}=v_{F}\tau_{z}\boldsymbol{\sigma}\cdot\mathbf{p}$
plus an interaction term with the same form in both valleys, i.e.,
$\tilde{\mathcal{V}}=\left(V_{0}+\Delta_{I}\sigma_{z}s_{z}\right)\Theta(R-r)$,
implying that gauge-invariant quantities, such as the polarization
and the conductivity, are insensitive to the choice of valley. Similar
arguments can be used in the case of Rashba-type scatterers or for
any type of SOC conserving the valley index.
\end{thebibliography}
\end{document}